\documentclass[apj]{emulateapj}
\usepackage{graphicx}
\usepackage{times}

\newcommand\approxlt{\mbox{$^{<}\hspace{-0.24cm}_{\sim}$}}
\newcommand{\be}{\begin{equation}}
\newcommand{\ee}{\end{equation}}
\newcommand{\bea}{\begin{eqnarray}}
\newcommand{\eea}{\end{eqnarray}}

\slugcomment{LBNL Report 61732}
\shorttitle{Improved Photometric Calibration of SDSS Imaging Data}
\shortauthors{Padmanabhan et al}

\begin{document}

\title{An Improved Photometric Calibration of the Sloan Digital Sky Survey Imaging Data}

\author{Nikhil Padmanabhan\altaffilmark{1,2,3}}
\email{NPadmanabhan@lbl.gov}
\author{David J. Schlegel\altaffilmark{1,4}}
\author{Douglas P. Finkbeiner\altaffilmark{4,5}}
\author{J. C. Barentine\altaffilmark{6,7}}
\author{Michael R. Blanton\altaffilmark{8}}
\author{Howard J. Brewington\altaffilmark{7}}
\author{James E. Gunn\altaffilmark{4}}
\author{Michael Harvanek\altaffilmark{7}}
\author{David W. Hogg\altaffilmark{8}}
\author{\v{Z}eljko Ivezi\'{c}\altaffilmark{9}}
\author{David Johnston\altaffilmark{10,4}}
\author{Stephen M. Kent\altaffilmark{11}}
\author{S.J. Kleinman\altaffilmark{12, 7}}
\author{Gillian R. Knapp\altaffilmark{4}}
\author{Jurek Krzesinski\altaffilmark{13,7}}
\author{Dan Long\altaffilmark{7}}
\author{Eric H. Neilsen, Jr.\altaffilmark{11, 7}}
\author{Atsuko Nitta\altaffilmark{14, 7}}
\author{Craig Loomis\altaffilmark{4, 7}}
\author{Robert H. Lupton\altaffilmark{4}}
\author{Sam Roweis\altaffilmark{15}}
\author{Stephanie A. Snedden\altaffilmark{7}}
\author{Michael A. Strauss\altaffilmark{4}}
\author{Douglas L. Tucker\altaffilmark{11}}

\altaffiltext{1}{Physics Division, Lawrence Berkeley National Laboratories, Berkeley, CA 94720-8160, USA}
\altaffiltext{2}{Joseph Henry Laboratories, Jadwin Hall, Princeton University, Princeton, NJ 08544, USA}
\altaffiltext{3}{Hubble Fellow, Chamberlain Fellow}
\altaffiltext{4}{Dept. of Astrophysical Sciences, Peyton Hall, Princeton University, Princeton, NJ 08544, USA}
\altaffiltext{5}{Harvard-Smithsonian Center for Astrophysics, 60 Garden St. Cambridge, MA 02138}
\altaffiltext{6}{The University of Texas, Department of Astronomy, 1 University Station, C1400, Austin, TX 78712-0259 USA}
\altaffiltext{7}{Apache Point Observatory, 2001 Apache Point Road, Sunspot, NM 88349, USA}
\altaffiltext{8}{Department of Physics, New York University, 4 Washington Pl, New York, NY 10003, USA}
\altaffiltext{9}{University of Washington, Department of Astronomy, Box 351580, Seattle, WA 98195, USA}
\altaffiltext{10}{Jet Propulsion Laboratory, 4800 Oak Grove Drive, Pasadena, CA 91109, USA}
\altaffiltext{11}{Fermi National Accelerator Laboratory, P.O. Box 500, Batavia, IL 60510, USA}
\altaffiltext{12}{Subaru Telescope, 650 N. A'ohoku Place, Hilo, HI 96720, USA}
\altaffiltext{13}{Mt. Suhora Observatory, Cracow Pedagogical University, ul. Podchorazych 2, 30-084 Cracow, Poland}
\altaffiltext{14}{Gemini Observatory, 670 N. A'ohoku Place, Hilo, HI 96720, USA}
\altaffiltext{15}{Dept. of Computer Science, University of Toronto, 6 King's College Rd., Toronto, Ontario, M5S 3G4, Canada}


\begin{abstract}
We present an algorithm to photometrically calibrate wide field 
optical imaging surveys, that simultaneously solves for the calibration parameters
and relative stellar fluxes using
overlapping observations. The algorithm decouples the problem of ``relative''
calibrations from that of ``absolute'' calibrations; the absolute calibration 
is reduced to determining a few numbers for the entire survey. We pay special 
attention to the spatial structure of the calibration errors, allowing one to 
isolate particular error modes in downstream analyses. Applying this to the Sloan
Digital Sky Survey imaging 
data, we achieve $\sim 1\%$ relative calibration errors across 8500 deg$^{2}$ in 
$griz$; the errors are $\sim 2\%$ for the $u$ band. These errors are 
dominated by unmodelled atmospheric variations at Apache Point Observatory. 
These calibrations, dubbed ``ubercalibration'', are now public with SDSS Data Release 6,
and will be a part of subsequent SDSS data releases.
\end{abstract}




\section{Introduction}
\label{sec:intro}

A common challenge for all physics experiments is relating a detector signal to the underlying
physical quantity of interest. Astronomical imaging surveys are no exception; 
a CCD camera counts
Analog to Digital units (ADU) in each pixel, a quantity that is
(approximately) proportional to the number of incident photons.
This relationship must be calibrated to yield physical
flux densities $(\rm{erg\, cm}^{-2} \rm{s}^{-1} \rm{Hz}^{-1})$. 
Key scientific programs of current and next generation imaging surveys
demand ever more precise photometric calibrations. For example, wide field imaging
surveys allow one to measure the clustering properties of
galaxies (and therefore, the underlying dark matter) on scales otherwise accessible only in
the Cosmic Microwave Background (CMB); comparing the CMB at
redshift $z \sim 1000$ with the relatively recent Universe at $z \approxlt 1$ allows
increasingly precise tests of our cosmological model. The first such measurements of
clustering on gigaparsec scales and larger were recently reported 
\citep{2007MNRAS.378..852P, 2007MNRAS.374.1527B}. These results emphasize the
need for accurate photometric calibrations over wide areas; the underlying clustering
signal is a rapidly decreasing function of scale, and could easily be overwhelmed by
percent-level systematic errors in the photometric calibration. A second example is
reconstructing the structure of the Galaxy, using the photometric properties of
different stellar populations. There have been a number of efforts to do this with
existing data \citep{2005astro.ph.10520J}, and it is a key scientific program for the next
generation of imaging surveys. Finally, there is the general (and powerful)
motivation that reducing systematic errors invariably reveals hitherto unseen details
and avenues of enquiry. Leveraging the current and next generation of imaging surveys
to yield their maximum scientific potential requires revisiting the problem of
photometric calibration, moving beyond the simplifications currently made
\citep{2006ApJ...646.1436S}. Several surveys are photometrically calibrated 
to a few percent; the challenge for the next generation of surveys
is to deliver $< 1\%$ calibrations over wide areas.

Photometric calibration involves relating the output of a 
CCD to the physical flux received above the Earth's atmosphere. 
For wide-field imaging surveys, 
we separate this into two orthogonal problems - ``relative'' calibration,
or the problem of establishing a consistent photometric calibration 
(albeit in possibly arbitrary flux units) across the entire
survey region, and ``absolute'' calibration, which transforms the relative calibrations 
into physical fluxes. This separation is useful since there exist a number of 
applications (such as the two discussed above) that are relatively 
tolerant of errors in the absolute calibration, but demand precise relative
calibrations. Current calibration techniques, which usually involve comparing 
observations to ``calibrated standards'', do not respect this distinction, making
it difficult to control errors in the relative calibration. Furthermore, calibrating
off standard systems normally involve relating different telescope
and filter systems, and are quickly limited by the accuracy with which these
transformations can be measured. Accurate relative calibrations would therefore
only use data from the native telescope/ filter system, obviating the need for 
any such transformations.

A second separation, emphasized by \cite{2006ApJ...646.1436S} 
is to separate the ``transfer function''
of the telescope and detectors from that of the atmosphere. The telescope and detectors
form an approximately closed system whose responses can be (potentially) mapped out with
exquisite precision with laboratory equipment. The atmosphere, on the other hand, is an open, 
highly dynamical, system with a range of relevant time scales; the best one can 
do is to monitor it with limited precision. Although we agree with this separation 
in principle, applying it would go significantly beyond the scope of this paper. We 
therefore do not make this distinction in the analysis presented here, but we return to 
it at the end of this work.

Techniques for relative photometric calibration have been
applied to optical imaging in the past, although much more
limited than the present work in the scope of either the number of objects or
the field of view.  \cite{1983AJ.....88..439L} and 
\cite{1992AJ....104..340L} are widely recognized as
describing one of the best-established ``photometric systems''.
Landolt observed several hundred stars near the celestial equator
with a photomultiplier tube on the Cerro Tololo 16-inch and the
Cerro Tololo 1.5-m telescopes.
Landolt achieved exceptional relative photometric calibrations
in five broad optical bandpasses (Johnson-Kron-Cousins UBVRI).
His data are accurate to 0.3\% per observation
of each star, with an even better accuracy implied for those stars
that have many observations.
Unfortunately, the full benefit of the accuracy of this
photometric system cannot be realized for other surveys due to the
systematic uncertainties in transforming from Landolt's system responses
to observations on other telescopes using (typically) CCD photometry.
There are some observations using exactly the Landolt system, most
famously of supernova 1987A, which made use of the otherwise-decommissioned
Cerro Tololo 16-inch telescope \citep{1987ApJ...320..589B}.

The other example of accurate relative optical photometry has come
from the searches for massive compact halo objects (MACHOs) from
microlensing events in dense star fields 
\citep[e.g.][]{1992AcA....42..253U,1993Natur.365..621A,1993Natur.365..623A}.  
MACHO events were detected from differencing images
taken with an identical instrument over timescales of several years.
More recently, these same techniques have been used to detect the
optical transits of planets.
With a proper treatment of the correlated noise properties in
the time series of images \citep{2006MNRAS.373..231P}, it is possible to
detect transits with peak depths of only $\sim$1\%. Note that the challenges 
here are different from the wide field imaging case considered in this work, 
since one is interested in differences in photometry of a single star.

Cosmic Microwave Background (CMB) anisotropy experiments also
demand very precise relative calibrations.
This accuracy is obtained with repeat 
observations of the sky, and cross-linked scan patterns. The redundancy thus obtained 
allows one to simultaneously solve for the CMB temperature at a given direction on the
sky and the detector calibration parameters. In this paper, we propose adopting this technique as a
new approach to calibrating optical imaging surveys - replacing the CMB temperature fluctuations
with the magnitudes of stars. Note that as this involves comparing multiple observations, 
this is a differential measurement, and therefore only yields a relative calibration. 
However, while the absolute calibration still must be obtained by comparison against standard 
stars, this is now applied uniformly across the entire survey region.
These ideas are not new to optical 
astronomy; precursors may be found in the work of \cite{1990MNRAS.246..433M,
1992PASP..104..435H,
1992MNRAS.257..650F, 1993A&A...271..714M, 1994IAUS..161..295F, 1994MNRAS.266...65G}. What is new 
to this work is both the (large angular) scales to which the method is applied 
and the accuracies obtained.

The Sloan Digital Sky Survey \citep[SDSS]{2000AJ....120.1579Y} is one of the most
ambitious optical imaging and spectroscopic
surveys undertaken to date. It has imaged a quarter of the sky in five optical bands, 
and has spectroscopically followed up more than a million of the detected objects. This 
makes the SDSS both a scientifically rich data set and an excellent proving 
ground for the next generation of surveys.
Accordingly, our goal in this paper is to develop the idea above in the context of the photometric
calibration of the Sloan Digital Sky Survey (SDSS). We begin by recapitulating 
aspects of the SDSS essential to this algorithm in Sec.~\ref{sec:SDSS}. Sec.~\ref{sec:algorithm}
then presents the details of the algorithm. We then assess the performance of our 
calibrations with simulations of the SDSS; the results are in Sec.~\ref{sec:simulations}. 
We then present a recalibration of the entire SDSS imaging data in Sec.~\ref{sec:sdssresults}.
Sec.~\ref{sec:public} announces the release of this calibration to the public, and
Sec.~\ref{sec:discussion} concludes with a discussion 
of the features and limitations of this work, as well as its
applicability to the next generation of imaging surveys. Although we focus on the SDSS, 
we phrase our discussion in terms that allow adapting the methods 
described here to arbitrary imaging surveys.

\section{The SDSS}
\label{sec:SDSS}

The Sloan Digital Sky Survey \citep{2000AJ....120.1579Y} is an
ongoing effort to image approximately a quarter of the sky, and
obtain spectra of approximately one million of the detected objects. The
imaging is carried out by drift-scanning the sky in photometric conditions
\citep{2001AJ....122.2129H}, using a 2.5m telescope \citep{2006AJ....131.2332G}
in five bands ($ugriz$)
\citep{1996AJ....111.1748F, 2002AJ....123.2121S} using a specially designed
wide-field camera \citep{1998AJ....116.3040G}.
These data are processed by completely automated pipelines that detect
and measure photometric properties of objects, and astrometrically
calibrate the data 
\citep{2001adass..10..269L, 2003AJ....125.1559P}. The
first phase of the SDSS is complete and has produced 
seven major data releases
\citep{2002AJ....123..485S, 2003AJ....126.2081A, 2004AJ....128..502A,
2005AJ....129.1755A, 2006ApJS..162...38A, 2007ApJS..172..634A,
2007arXiv0707.3413A}\footnote{\texttt{http://www.sdss.org/dr6}}.

\begin{figure}
\begin{center}
\leavevmode
\includegraphics[width=3.0in,angle=90]{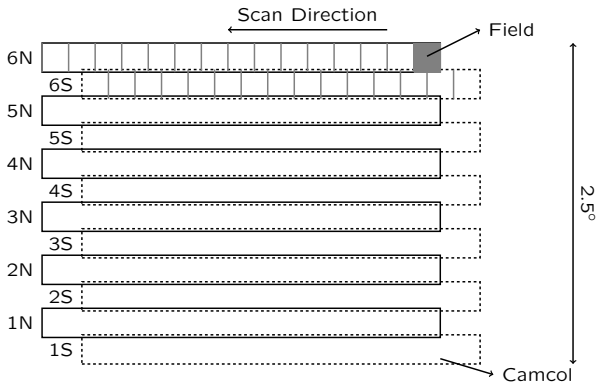}
\end{center}
\caption{The geometry of the SDSS imaging. Part of an SDSS stripe with the two
interleaved strips (denoted by N and S) is shown. Each strip consists of six camcols (numbered
1 through 6 in the figure), while each camcol is further divided into fields (for simplicity, 
we show field divisions for only two camcols). See the text for more details.}
\label{fig:geom}
\end{figure}

The SDSS imaging data (see also Fig.~\ref{fig:geom}) 
are taken by drift-scanning along {\it stripes} centered on 
great circles on the sky in all five filters. These stripes are $2.5^{\circ}$ wide, and are
filled by two interleaved {\bf \it strips}. The actual data is taken in {\it runs},
which are part of strips; multiple runs may be taken in a single night, not necessarily
on the same strip. Each run is further subdivided into six {\it camera columns} or 
{\it camcols}, corresponding to the six columns of CCDs on the camera. The data from
each CCD is in turn split into {\it frames}, consisting of 1361 drift scan rows.
The five frames corresponding to the same region of sky observed in the five SDSS filters,
are collectively referred to as a {\it field}.
Note that while runs and camcols correspond to physical separations of the data, the division
into frames is purely artificial. The integration time is approximately 54.1 seconds per frame
in each filter, with a time lag of $\sim 73$ seconds between each adjacent filter. The order
of the filters as they observe the sky is $riuzg$.

\begin{figure*}
\begin{center}
\leavevmode
\includegraphics[width=6.0in]{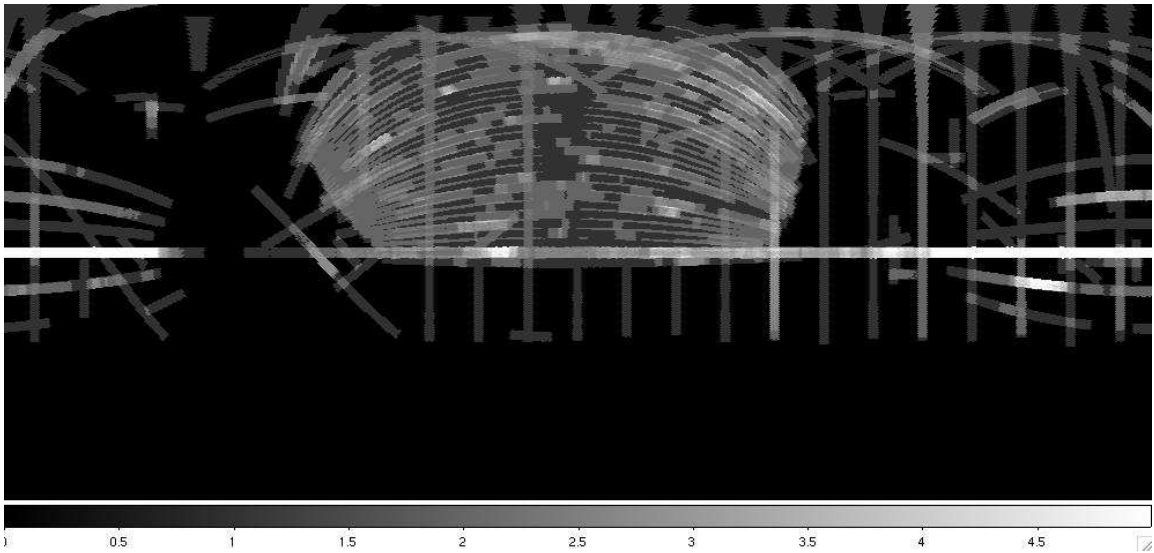}
\end{center}
\caption{The sky coverage of the SDSS data used in this paper, shown 
in an equal area resolution 7 HEALPIX/HEALCART 
\citep{1999elss.conf...37G,2004ApJ...614..186F} projection. 
The $x$ scale covers RA $0^{\circ}$ to $360^{\circ}$, while the 
$y$ axis runs from DEC $90^{\circ}$ to $-90^{\circ}$. The grey scale denotes
the mean number of observations of a star in a particular pixel. Note that we
saturate at 5 observations, although on the equatorial (white) stripe, there are 
pixels with a mean number of observations as high as 15. The bulk of the survey
data is in the North Galactic Cap, the prominent structure in the center of the 
image. The Equatorial stripe, imaged every Fall, is the white horizontal stripe
halfway in the image. The approximately equally spaced vertical runs are examples of
the Apache Wheel data.}
\label{fig:sky_coverage}
\end{figure*}

The current survey flux calibrations are applied in a three step process, 
involving three different telescopes and subtly different filter systems. The absolute flux
system is defined by  BD+17$^{\circ}$4708 \citep{1983ApJ...266..713O}, 
an F0 subdwarf star and is based on synthetic photometry in the expected (at the time)
SDSS $u'g'r'i'z'$ filters and an improved spectral energy distribution for this star 
\citep{1996AJ....111.1748F}. This is used to calibrate a primary network
of 158 stars observed by the USNO 40-inch at Flagstaff in Arizona, chosen to span a range
in color, airmass, and RA, and distributed over the Northern sky \citep{2002AJ....123.2121S}. 
Unfortunately, these stars
saturate the SDSS survey telescope; the calibrations are therefore indirectly transferred via
a 20 inch Photometric Telescope (PT) at Apache Point (APO), which observes these primary 
standards, as well as 1520 $41.5 \times 41.5\,{\rm arcmin}^{2}$ {\it secondary patches} of sky.
These patches are what finally calibrate the data from the 2.5m survey 
telescope \citep{2006AN....327..821T}. Note that these are three different filter
systems, and not just realizations of one system. 
In addition to conflating the absolute and relative calibrations, 
this indirect transfer of the calibration makes 
achieving $< 1\%$ calibrations via this method a challenge, although it does return 
relative calibrations accurate to $\sim 2\%$ \citep{2004AN....325..583I}. 
Note that these errors have natural scales
of $2.5^{\circ}/12$ (the width of a camcol) and $2.5^{\circ}$ (the width of a stripe) 
perpendicular to the scan direction. 

Before continuing, it is worth emphasizing that $2\%$ calibrations for
a wide field optical survey was unprecedented until very recently.
However, motivated by the promise of future wide field surveys and the 
challenge of $1\%$ photometry, we realized 
that the next step {\it must} short-circuit the above multi-stage calibration
pipeline.
The calibration algorithm we propose here relies on repeat observations to 
constrain the photometric calibrations. Unfortunately, in the standard
survey strategy, the only significant repeat observations occur at the poles of the survey
(where the great circles of stripes converge), and on the celestial equator, which is
re-imaged every Fall (see Fig.~\ref{fig:sky_coverage}). 
While the Fall equatorial stripe has sufficient repeat observations
to make precise photometry possible \citep{2007astro.ph..3157I}, the calibration for the
bulk of the survey region would only be constrained at the survey poles, clearly
undesirable. The only other natural overlaps occur when 
the beginning and ends of runs overlap each other along strips. While
this does connect the survey from one pole to the other, most of the overlaps occur on 
the same CCD column, and so have limited utility since these are degenerate with 
flat fields.

To address both these inadequacies, two additional sets of data were taken. The first 
were short scans that cross the normal scan directions. Such {\it oblique} scans 
exist for most observing years (Fall through Spring), and 
were taken to check for temporal variations of the flat fields. These are invaluable for constraining
flat fields, since they compare each CCD column with every other. The other
data were a grid of long scans, dubbed the ``Apache Wheel'', designed to connect 
every part of the survey with every other. Observing
such a grid with the usual SDSS scanning speed would have required a significant
expenditure of telescope time, adversely affecting the science goals of the survey.
The compromise was to observe these data, at 7 times the normal scanning speed 
(i.e. with an effective exposure time of $\sim 8$ seconds), and 
binning data into 4x4 native camera pixels. Reducing these data required modifications to the 
survey data reduction pipelines \citep{2001adass..10..269L}, and was done at Princeton
(along with a re-reduction of regular survey data)
as part of this calibration effort.
The survey region we consider in this paper is in Fig.~\ref{fig:sky_coverage}, with the 
greyscale encoding the number of repeat observations of different regions of the sky.

\section{The Algorithm}
\label{sec:algorithm}

\subsection{The Photometric Model}
\label{sub:photometric_model}

An introduction to photometric calibrations and photometric standard systems may
be found in \cite{2005ARA&A..43..293B}; we focus on the details relevant to this work 
below.
Assuming linearity, the flux $f$ of an object at Earth (above the atmosphere) is 
related to the detected flux $f_{ADU}$\footnote{An ADU, or
Analog-Digital Unit is the digitization of the analog detector output} by
\begin{equation}
\label{eq:photometric1}
f = {\cal K} f_{ADU} \,\,;
\end{equation}
the problem of photometric calibration is to determine ${\cal K}$.
The above equation is deceptively compact; ${\cal K}$ depends on the 
exposure time, detector efficiency, filter responses, 
the telescope optical system, the optical path through the atmosphere, 
the spectral energy distribution of the objects in question, and all
the variables that these in turn are sensitive to. Furthermore, Eq.~\ref{eq:photometric1}
makes no reference to the units of $f$ and ${\cal K}$; the problem of determining the 
correct units is that of {\it absolute} photometric calibration; we restrict our
discussion below to the problem of {\it relative} calibrations.

Since all the above terms affect the flux multiplicatively, it is convenient
to work in log-space; the above effects become additive corrections.
Converting fluxes to magnitudes ($m = -2.5 \log_{10} f$), Eq.~\ref{eq:photometric1} becomes
\begin{equation}
\label{eq:photometric2}
m = m_{ADU} - 2.5 \log_{10} ({\cal K}) \,\,.
\end{equation}
Expanding ${\cal K}$ in terms of its various dependencies, we obtain
\begin{equation}
\label{eq:calk}
-2.5 \log_{10} ({\cal K}) = a(t) - k(t)x + f(i,j;t) + ... \,\,,
\end{equation}
where all terms are a function of time. The optical response of the telescope
and detectors is the ``a-term'' $a(t)$, while the detector flat fields (in magnitudes)
are $f(i,j;t)$
where $i,j$ represent CCD coordinates.
The atmospheric extinction is the product of the ``k-term'' $k(t)$ and
the airmass of the observation, $x$. Note that this is a crude 
phenomenological model
(it heuristically resembles a first order Taylor expansion), but is completely
adequate for our purposes. We therefore defer a discussion of its limitations
and potential extensions to Sec.~\ref{sec:discussion}.

\begin{table}
\caption{\label{tab:flatseasons} Flat Field seasons}
\begin{tabular}{cccl}
\tableline
SDSS Run & MJD  & Date &  Comments \\
\tableline
    1 & 51075  & 19-Sep-1998  &  Beginning of Survey \\
  205 & 51115  & 28-Oct-1998  &  \\
  725 & 51251  & 13-Mar-1999  &  \\
  941 & 51433  & 12-Sep-1999  &  \\
 1231 & 51606  & 03-Mar-2000  &  \\
 1659 & 51790  & 03-Sep-2000  & After i2 gain change \\
 1869 & 51865  & 17-Nov-2000  & Vacuum leak in Dec 2000 \\
 2121 & 51960  & 20-Feb-2001  & After vacuum fixed \\
 2166 & 51980  & 12-Mar-2001  & \\
 2504 & 52144  & 23-Aug-2001  & After summer shutdown \\
 3311 & 52516  & 30-Aug-2002  & After summer shutdown \\
 4069 & 52872  & 20-Aug-2003  & After summer shutdown \\
 4792 & 53243  & 26-Aug-2004  & After summer shutdown \\
 5528 & 53609  & 26-Aug-2005  & After summer shutdown \\
\tableline
\end{tabular}
\tablecomments{The starting dates, and the corresponding
first SDSS run for the flat field seasons.}
\end{table}

We now specialize to the SDSS;
we calibrate each of the five filters individually, and 
assume that each of the six camera columns are independent, yielding 
an a-term and flat field to be determined per CCD.
We implicitly assume that the filter response for each of the six
CCDs is identical (we return to this in Sec.~\ref{sec:discussion}).
The k-term is however common to all camera columns and 
depends only on the filter. Also, since the SDSS observes by 
drift-scanning the sky, the flat fields are no longer two-dimensional, but only
depend on the CCD column and are represented by a 2048 element vector. This 
is complicated by the fact that some of the SDSS CCDs have two amplifiers, resulting
in a discontinuity at the center of the flat field. 
We model this by assuming the flat fields have the form,
\begin{equation}
\label{eq:flatfieldeq}
f(i,j)= f_{0}(j) + \theta(j-1024)\Delta f
\end{equation}
where $\theta(x)$ is the Heaviside function, and $\Delta f$ (hereafter, the ``amp-jump'')
is the relative gain of the two amplifiers. 
Note that as written, $f_{0}$ is a continuous function of CCD column.
Finally, we need to specify the time dependence 
of these quantities. The a-terms and amp-jumps are assumed to be constant during a night,
and we simply specify these as piecewise constant functions.

It was also realized early in the survey (about 2001) that the flat fields were time-dependent,
and appeared to be changing discontinuously over the summers when the camera was disassembled 
for maintenance. These changes are most likely associated with changes in the surface chemistry
of the CCDs. We therefore model the flat fields as being constant in time over a ``flat field season'', 
roughly the period between any maintenance of the camera. 
The boundaries, in MJD and SDSS run number, of these ``seasons'' 
are listed in Table~\ref{tab:flatseasons}.
Ideally,
one might have chosen an even finer time interval to test the constancy of the flat fields; 
however, the SDSS lacks sufficient oblique scan data to improve the time resolution. We note here
that the standard practice of measuring flat fields from sky data does not work for the SDSS, due
to scattered light in the camera.

The time dependence of the k-terms at APO is more complicated, as the atmosphere (on average) gets
more transparent as the night progresses, at the rate of $\sim 1$ mmag/hour (millimagnitudes/hour) per 
unit airmass. We therefore model $k(t)$ over the course of a night
as 
\begin{equation}
\label{eq:kterm_def}
k(t) = k + \frac{dk}{dt} (t - t_{ref}) \,\,,
\end{equation} 
where $t_{ref}$ is a reference time\footnote{We adopt 0700 UT as $t_{ref}$, corresponding to
midnight Mountain Standard Time.}. Note that $t$ in the above equation only runs over the course of a single
night; $k$ and $dk/dt$ can (in principle) vary from night to night, and there is no
requirement on the continuity of $k(t)$ across nights. Table~\ref{tab:paramtable} 
summarizes the parameters in our photometric model, whose final form is 
\begin{equation}
\label{eq:photo_eqn}
m = m_{ADU} + a_{\alpha} - \left[k_{\beta} + 
\left(\frac{dk}{dt}\right)_{\beta}(t - t_{\beta,ref})\right] x
 + f_{\gamma}(j) \,,
\end{equation}
with $\alpha$, $\beta$, and $\gamma$ indexing the appropriate a-term, k-term (and $t_{ref}$), 
and flat field for the star in question.

\begin{table}
\caption{\label{tab:paramtable} Calibration Parameters}
\begin{tabular}{lccl}
\hline
Parameter & Number & Fit & Comments \\
\hline
a-terms & $6 \times 5 \times n_{night}$  & Yes & \\
$k$ & $5 \times n_{night}$ & Yes & k-term at $t=t_{ref}$ \\
$dk/dt$ & $5$ & No & \\
Flatfields & $6 \times 5 \times n_{season}$ & Yes & 2048 element vector \\
& & (iterative) & \\
Ampjumps & $6 \times 5 \times n_{run}$ & No & \\
\hline
\end{tabular}
\tablecomments{The parameters that make up the photometric
model. The number of parameters are functions of $n_{night}$ (the number of nights),
$n_{season}$ (the number of flatfield seasons), $n_{run}$ (the number of runs), and 
the number of filters (5), and camera columns (6). Also listed is whether the parameter 
is fit for or not.}
\end{table}

\subsection{Solution}
\label{sub:solution}

Having specified the parameters of the photometric model, we now turn to the problem 
of determining them. It is natural 
to consider repeat observations of stars to constrain these parameters\footnote{While
in principle, one could also consider galaxies, we restrict our discussion to stars 
to avoid subtleties of extended object photometry.}. Let us therefore consider $n_{obs}$
observations with observed instrumental magnitudes $m_{ADU,j}$,
of $n_{star}$ unique stars with unknown true magnitudes $m_{i}$. 
Note that $n_{obs}$ is the number of observations of {\it all} stars, i.e. 
$n_{obs} = \sum_{i=1}^{n_{star}} n_{i}$ where $n_{i}$ is the number of times star $i$ is
observed.
Using Eq.~\ref{eq:photo_eqn},
we construct a $\chi^{2}$ likelihood function for the unknown magnitudes and photometric
parameters,
\begin{equation}
\label{eq:photo_chi2tot}
\chi^{2}[a_{\alpha}, k_{\beta}, (dk/dt)_{\beta}, f_{\gamma}]
 = \sum_{i}^{n_{star}} \chi^{2}_{i} \,\,,
\end{equation}
with 
\begin{equation}
\label{eq:photo_chi2i}
\chi^{2}_{i} = \sum_{j \in {\cal O}(i)} \left[\frac{m_{i} - m_{j,ADU} - a_{\alpha(j)} +
k_{\beta(j)}(t)x - f_{\gamma(j)}}{\sigma_{j}}
\right]^{2} \,,
\end{equation}
where $j$ runs over the multiple observations, ${\cal O}(i)$, of the $i^{th}$ star, 
$\sigma$ is the error in $m_{j,ADU}$, and $k(t)$ is given by Eq.~\ref{eq:kterm_def}. 
We also assume that errors in observations are independent; this is 
not strictly true as atmospheric fluctuations temporally correlate different observations.
One can generalize the above to take these correlations into account and,
as we show below, our results are not biased by this assumption.
Note that Eq.~\ref{eq:photo_chi2tot} has $n_{obs}$ known quantities, and $n_{star} + \#({\rm
parameters})$ unknowns. In general, the number of photometric parameters $\ll n_{star}$, 
and $n_{obs} > 2n_{star}$, implying that this is an overdetermined system. 

To proceed, we start by minimizing Eq.~\ref{eq:photo_chi2tot} with respect to $m_{i}$; this 
yields,
\begin{eqnarray}
\label{eq:min_chi2_m}
\frac{d\chi^{2}}{dm_{i}} & = 2 \sum_{j \in {\cal O}(i)} \left[\frac{m_{i} - m_{j,ADU} - a_{\alpha(j)} +
k_{\beta(j)}(t)x - f_{\gamma(j)}}{\sigma_{j}^{2}}\right] \nonumber \\
& = 0 \,,
\end{eqnarray}
which is trivially solved for $m_{i}$ to give,
\begin{eqnarray}
\label{eq:min_chi2_m1}
m_{i} = \sum_{j \in {\cal O}(i)} \left[\frac{m_{j,ADU} + a_{\alpha(j)} -
k_{\beta(j)}(t)x + f_{\gamma(j)}}{\sigma_{j}^{2}}\right] \nonumber \\
\times \left[\sum_{j \in {\cal O}(i)} \left(\frac{1}{\sigma_{j}^{2}}\right)\right]^{-1} \,.
\end{eqnarray}
As substituting the above result into Eq.~\ref{eq:min_chi2_m} to solve for the 
calibration parameters is 
algebraically unwieldy, we reorganize these results by making
the following notational change. We arrange the unknown photometric parameters into an 
$n_{par}$ element vector ${\bf p}$,
\begin{equation}
\label{eq:xdefn}
{\bf p} = \left(
\begin{tabular}{c}
$a_{\alpha}$ \\
$k_{\beta}$ \\
$(dk/dt)_{\beta}$ \\
$f_{\gamma}$ \\
\end{tabular}
\right) \,\,.
\end{equation}
Then substituting Eq.~\ref{eq:min_chi2_m1} into Eq.~\ref{eq:photo_chi2i} yields a matrix
equation for $\chi^{2}$,
\begin{equation}
\label{eq:chi2_matrix}
\chi^{2} = ({\bf Ap-b})^{t} {\bf C}^{-1} ({\bf Ap-b}) \,\,,
\end{equation}
where ${\bf A}$ is an $n_{obs} \times n_{par}$ matrix , 
and ${\bf b}$ is an $n_{obs}$ element vector, and ${\bf v}^{t}$ represents
the transpose of ${\bf v}$. The 
errors are in the covariance matrix ${\bf C}$ which, in Eq.~\ref{eq:photo_chi2i},
is assumed to be diagonal (but can be generalized to include correlations between
different observations). For clarity, we explicitly write out the form of 
${\bf Ap - b}$ for the case of a single star observed twice at airmass
$x_{1}$ and $x_{2}$, and with errors
$\sigma_{1}$ and $\sigma_{2}$, where only the 
a- and k-terms are unknown,
\begin{eqnarray}
\left[
\left(
\begin{tabular}{cccc}
1 & 0 & $-x_{1}$ & 0 \\
0 & 1 & 0 & $-x_{2}$ \\
\end{tabular}
\right)
-\left(
\begin{tabular}{cccc}
$I_{1}$ & $I_{2}$ & $-x_{1}I_{1}$ & $-x_{2}I_{2}$ \\
$I_{1}$ & $I_{2}$ & $-x_{1}I_{1}$ & $-x_{2}I_{2}$ \\
\end{tabular}
\right)
\right] \nonumber 
\end{eqnarray}
\begin{eqnarray}
\label{eq:eq:chi2_matrix_explicit}
& & \times
\left(
\begin{tabular}{c}
$a_{1}$ \\
$a_{2}$ \\
$k_{1}$ \\
$k_{2}$ \\    
\end{tabular}
\right) \nonumber \\
& & - 
\left(
\begin{tabular}{c}
$m_{1,ADU} - m_{1,ADU}I_{1} - m_{2,ADU}I_{2} $ \\
$m_{2,ADU} - m_{1,ADU}I_{1} - m_{2,ADU}I_{2} $\\
\end{tabular}
\right) \,\,,
\end{eqnarray}
where $I_{i}$ is the normalized inverse variance, 
$I_{i} = \sigma_{i}^{-2}/\sum_{j} \sigma_{j}^{-2}$. 
Each row of ${\bf Ap-b}$ has a simple interpretation as the difference
between the magnitude of a particular observation of a star, and the inverse
variance weighted mean magnitude of all observations of that star. Also, 
although ${\bf A}$ is a large matrix
($\sim 50,000,000 \times 2000$ for the SDSS), it is extremely sparse, and
amenable to sparse matrix techniques.

Obtaining the best fit photometric parameters simply involves minimizing
Eq.~\ref{eq:chi2_matrix}. Although there are several choices here, we proceed
via the normal equations \citep[e.g.,][]{1992nrfa.book.....P},
\begin{equation}
\label{eq:chi2_normal}
\frac{d\chi^{2}}{d{\bf p}} = {\bf A}^{t} {\bf C}^{-1} {\bf Ap} - 
{\bf A}^{t} {\bf C}^{-1} {\bf b} = 0 \,\,.
\end{equation}
The inverse curvature matrix,
\begin{equation}
\label{eq:chi2_curve}
\frac{d^{2}\chi^{2}}{dp_{i}dp_{j}} = \left({\bf A}^{t} {\bf C}^{-1} {\bf A}\right)_{ij} \,\,,
\end{equation}
provides an estimate of the uncertainty in the recovered parameters. Note that it is however
{\it not} the covariance matrix of the parameters, since Eq.~\ref{eq:chi2_matrix} was derived
marginalizing over the unknown magnitudes of all the stars. Furthermore, since the 
measurement errors do not account for temporal variations in the atmosphere, the ``error''
estimates from the curvature matrix may be significantly underestimated.

We conclude by noting the similarities between the above and algorithms used for 
making maps of the CMB \citep[e.g.][]{1997ApJ...480L..87T}
\footnote{This is not accidental, 
as this work was inspired by the techniques learned in CMB mapmaking.}. The SDSS
runs are analogous to CMB scan patterns, while the magnitudes are equivalent to the
temperature measurements. However, unlike the CMB, our principal goal is the 
calibration parameters,
with the magnitudes of the stars being a secondary product\footnote{This results in the 
unusual situation of having $\sim$ a few million nuisance parameters that must 
be marginalized over to obtain $\sim$ a thousand parameters of interest.}.

\subsection{Degeneracies and Priors}
\label{sub:degeneracies_and_priors}

Our choice of photometric parameters is non-minimal in that there exist degeneracies
between them. These degeneracies are of more than academic interest, as they 
make the normal equations singular, and solutions of them unstable. 
We now  discuss the source of these degeneracies, and how the resulting numerical instabilities can 
be tamed by the use of priors.
\begin{itemize}
\item {\it Zero point} : As the above algorithm is based solely on magnitude differences, 
any overall additive shift of all the a-terms does not change $\chi^{2}$. Note that this 
is simply the problem of absolute calibration rephrased. 
\item {\it Disconnected Regions} : This is a generalization of the previous case; the zero
points of each disconnected region of the survey can be individually changed, without changing
$\chi^{2}$. Note that disconnected in this context refers regions with 
neither spatial nor temporal overlap (as we assume the photometric parameters are stable
over the course of a night) with other parts of the survey.
\item {\it Zero point of flats} : In Eq.~\ref{eq:photo_eqn}, 
the zero point of the flat fields
is degenerate with the a-terms; this degeneracy is trivially lifted by forcing the 
flat fields to have zero mean.
\item {\it Constant Airmass} : The photometric equation schematically is $\sim a - kx$; 
therefore for data with little or no airmass variation, there is a degeneracy direction that
keeps $a-kx$ constant, while changing both the a- and k-terms. While this does not affect 
the calibration in regions where $a-kx$ is constrained, extrapolating the a- and k-terms to regions 
with different airmasses can result in incorrect calibrations.
\end{itemize}

There is a useful generalization of the above discussion; the inverse eigenvalues of the curvature
matrix (Eq.~\ref{eq:chi2_normal} and \ref{eq:chi2_curve}) are a measure of the error on the 
determination of linear combinations of the photometric parameters (encoded by the corresponding 
eigenvectors). The degeneracies discussed above are characterized 
by eigenvalues $\sim 0$, which make the normal equations 
unstable. However, any badly constrained combinations 
(even if they are {\it formally} well determined)
can amplify noise and un-modeled systematics in the data, potentially introducing errors 
when the calibrations are applied. We therefore identify all eigenvectors of photometric parameters
that are poorly constrained
(i.e. those that could result in potential errors of $>$ 1\% ), and project these out; 
this renders the normal equations stable 
and they can be directly solved. Note that this introduces a tunable parameter to the solution - the 
eigenvalue threshold below which we project out modes. This threshold is chosen such 
that the final calibrations are insensitive to its exact value.

Although projecting out poorly determined eigenvectors yields a minimal set of parameters well 
constrained by the data, we must add back in these ``null'' eigenvectors to get a solution in 
our original (and preferred) parameter space. We achieve this by introducing priors on the photometric
parameters, and then adjusting the values of all the photometric parameters 
along the null vectors to best satisfy these priors. Assuming equally weighted Gaussian priors
on the parameters ${\bf p}$
with a mean value ${\bf p}_{0}$, this can be phrased as an auxiliary $\chi^{2}$ minimization,
\begin{equation}
\label{eq:chi2_prior}
\chi^{2}_{\rm prior} = |{\bf \hat{p}'} + {\bf V}_{null} \delta{\bf x} - {\bf p}_{0}|^{2} \,\,,
\end{equation}
where ${\bf \hat{p}'}$ is the solution of the normal equations, Eq.~\ref{eq:chi2_normal}, and
we have gathered the $n_{degen}$ null eigenvectors 
into a $n_{par} \times n_{degen}$ matrix ${\bf V}_{null}$. Varying 
$\delta{\bf x}$ to minimize $\chi^{2}_{\rm prior}$, we obtain our final solution for 
the photometric parameters, 
\begin{equation}
\label{eq:chi2_soln}
{\bf \hat{p}} = {\bf \hat{p}'} + {\bf V}_{null} \delta{\bf x} \,\,.
\end{equation}

\subsection{Implementation Details}
\label{sub:implement}

The above discussion described our calibration algorithm in generic terms, with minimal
reference to survey specifics. We now discuss the details and approximations specific to 
implementing this algorithm for the SDSS.

The first approximation involves determining the flat field vectors. As described, the flat
field vectors are determined simultaneously with the other photometric parameters. Doing so
would however have approximately doubled the number of photometric parameters, and significantly
complicated the degeneracies between the various parameters. We therefore chose an iterative 
scheme where the flat fields are held constant while the other parameters are determined. We 
then use the best fit solution to measure the magnitude differences between multiple 
observations as a function of CCD column, and fit a flat field vector to these via a 
quadratic B-spline with 17 uniformly spaced knots. 
As we show in the next section, this scheme rapidly converges to the true solution.
In addition, the SDSS photometric pipeline estimates the amp-jumps by requiring that the background
be continuous across the amplifiers. 
Instead of fitting to the amp-jumps, we simply hold them fixed to these values.

The second approximation involves the k-terms and their time derivatives. The 
typical airmass variations over the course of a single night tend to be small, making the determination
of k-terms very degenerate with the a-terms, as discussed in the previous section. The situation
is even more degenerate for the time derivative of the k-terms. We fix these degeneracies by using 
priors for the k-terms, and fixing their time derivatives to
values estimated by the SDSS photometric telescope \citep{2001AJ....122.2129H}.
Table~\ref{tab:paramtable} summarizes 
which parameters are fit in our implementation, while Table~\ref{tab:maglimit} lists the 
mean values for $k$ and $dk/dt$.

We must also specify the actual objects used for calibrating. We restrict ourselves to objects that the
SDSS classifies as stars, and use aperture (7.43 arcsecond radius) 
photometry to determine their magnitudes. The first choice sidesteps 
the subtleties of galaxy photometry, while aperture photometry avoids aliasing errors from the point 
spread function (PSF) estimation into the calibration.  
The magnitude limits we use are in Table~\ref{tab:maglimit}, 
along with the number of unique stars,
and observations. We choose not to make any color cuts on the stars to eliminate
variable stars and quasars. These only add noise to any calibrations, but cannot bias the results;
we therefore just use outlier rejection ($3\sigma$ clipping) 
and iterate our algorithm to minimize such contamination. 
A significant advantage of this approach is that the calibration of the 5 SDSS filters is
independent, allowing us to use colors of sub-populations of stars as external tests of the calibrations;
this is discussed in detail in Sec.~\ref{sub:principal_colors}.

Our algorithm does assume that the input data were taken under photometric conditions. We therefore,
at the outset, 
eliminate all data taken under manifestly non-photometric conditions. As we discuss below, the algorithm
does provide diagnostics of the photometricity of the data; we therefore iterate the algorithm removing
any remaining non-photometric data.

\begin{table}
\caption{\label{tab:maglimit}}
\begin{tabular}{ccccccc}
\hline
Filter & Mag. & $n_{star}$ & $n_{obs}$ 
& $k_{0}$ & $dk/dt $ 
& $\sigma(dk/dt)$\\
& Limit & $(\times 10^{6})$ & $(\times 10^{6})$ & &  $(\times 10^{2})$ & $(\times 10^{2})$ \\
\hline
$u$ & 18.5 &  4.7 & 14.6 & 0.49 & - 1.2 & 2.5 \\
$g$ & 18.5 &  9.3 & 29.1 & 0.17 & -0.7 & 1.7 \\
$r$ & 18.0 & 11.7 & 36.5 & 0.10 & -1.0 & 1.7 \\
$i$ & 17.5 & 11.5 & 35.9 & 0.06 & -1.2 & 1.5 \\
$z$ & 17.0 & 11.6 & 36.4 & 0.06 & -2.2 & 1.7 \\
\hline
\end{tabular}
\tablecomments{The magnitude limits used to select stars for calibration
for the 5 SDSS filters, with the resulting number of unique stars, $n_{star}$, and the 
total number of observations, $n_{obs}$ (in millions of stars). 
Also listed are the mean k-term $k_{0}$ (used as a prior),
the mean time variation of the 
k-term, $dk/dt$ (in mags/airmass/10 hours), and its scatter about the mean. The latter
is used in our simulations to determine the step size for the random walk approximation
to the atmospheric extinction. Note that we do not fit for the time variation of the 
k-term but simply use the values for the entire survey.}
\end{table}

Finally, there remains the issue of the absolute calibration of these data, or determining 
the five zero points for each filter. Improving the absolute calibration is beyond the 
scope of this paper; we therefore determine the zero points by matching magnitudes on average
to those obtained by the standard SDSS calibration pipeline. These are therefore essentially on 
an AB system \citep{2004AJ....128..502A}, tied to the SDSS fundamental 
standard, BD+17$^{\circ}$4708 \citep{1983ApJ...266..713O}.

\section{Simulations}
\label{sec:simulations}

Simulations serve the dual purpose of verifying the above algorithm and 
our implementation of it, as well as quantifying the level of 
residual systematics. We construct the simulations as follows :
\begin{itemize}
\item We start with the actual catalog of stars observed by the SDSS, 
with the magnitude cuts described above. This ensures (by construction)
that the pattern of overlaps in the simulations matches the observed data,
essential to obtaining realistic results.
\item We simulate ``true'' magnitudes for each of the stars, using a
power law distribution, where the normalization and slope are matched to 
their observed values.
\item Given an observation of the star, we then transform the magnitude 
into an observed instrumental magnitude, assuming values for the a- and k-terms
and flat fields.
We simulate the time variation of the k-term by describing $k(t)-k_{0}$
by a Gaussian random walk in time, with a drift in time given by $dk/dt$. The size of 
the steps
is set by the observed value of the scatter in $dk/dt$ (Table~\ref{tab:maglimit}). 
Note that this random
component attempts to model the correlations in time induced by the 
atmosphere, albeit by making the simplification that the spectrum of fluctuations
is described by a Gaussian random walk. Non-photometric data is 
simulated by exactly the same process, although we arbitrarily increase the 
scatter in the random walk.
\item We add noise to the instrumental magnitudes by considering the 
Poisson noise from both the object and the sky. 
Note that the Poisson fluctuations from the sky dominate the error budget
for most of the objects.
\end{itemize}
These simulated catalogs are structurally identical to the actual data. We can therefore
analyze them in {\it exactly} the same manner, and compare the derived parameters
with those input, providing us with an end-to-end test of our pipeline. Furthermore, 
these simulations have exactly the same footprints, timestamps and overlap patterns as
the real data, allowing us to estimate our final errors and explore parameter degeneracies.
 
\subsection{Results}
\label{sub:simresults}

\begin{figure}
\begin{center}
\leavevmode
\includegraphics[width=3.0in]{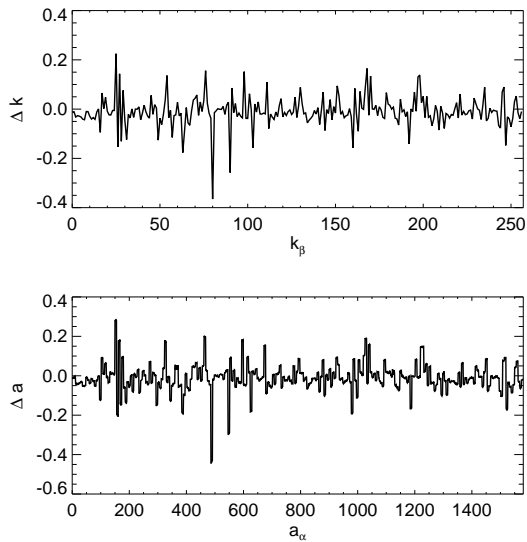}
\end{center}
\caption{The difference between the estimated and true a- and k-term for the
$r$ band in one of our simulations. There are approximately 6 a-terms that correspond to a given k-term,
and the scales on the $x$ axis are adjusted so that corresponding terms are aligned.
Note that the estimated a- and k-terms are highly covariant.
}
\label{fig:dadk}
\end{figure}

\begin{figure}
\begin{center}
\leavevmode
\includegraphics[width=3.0in]{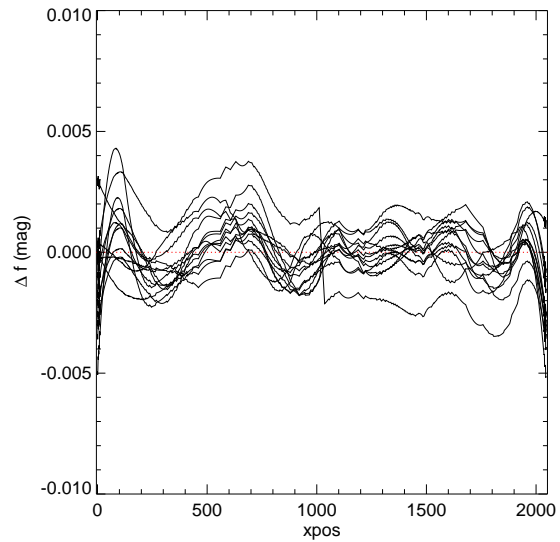}
\end{center}
\caption{The difference between the estimated and true flat field vectors 
for the $r$ band of one of our simulations. 
Each line corresponds to a different flat field season.
Since the mean of the flat fields are 
degenerate with the a-terms, we only plot the deviations about the mean.
For clarity, only the flat field vectors for one camera column
are plotted; the results for the other camera columns are similar. The errors
in the flat field estimation are $\sim 0.5\%$ (peak to peak).
}
\label{fig:sim_dflat}
\end{figure}

Figs.~\ref{fig:dadk} and \ref{fig:sim_dflat} show the differences between the true
and estimated a- and k-terms, and flat field vectors for the $r$ band in one of our
simulations, analyzed identically to the real SDSS data. The flat field vectors are
recovered with an error $< 0.5\%$. 
The SDSS pipeline stores flat fields as scaled integer arrays; the
roundoff error from this is about an order of magnitude lower.
The a- and k-terms are similarly correctly
estimated on average, although there are significant misestimates for both. However, 
a striking feature of Fig.~\ref{fig:dadk} is the similarity in the residuals for the 
a- and k-terms, reminiscent of the discussion of the degeneracies between the a- and k-terms
in Sec.~\ref{sub:degeneracies_and_priors}. This suggests comparing the estimated and true
values of $a-k\langle x\rangle$ on a per-field basis, where $\langle x\rangle$ is the average
airmass over a given field and filter; it is this combination that determines the photometric
calibration of a field.

\begin{table}
\caption{\label{tab:caliberr} Calibration Errors}
\begin{center}
\begin{tabular}{cccccc}
\hline
Filter & $\langle \Delta m \rangle$ & $\sigma$ & $\sigma_{3}$ & $\%(3\sigma)$ & $\sigma_{0}$ \\
\hline
$u$& -1.67& 13.38& 12.53&  0.85&  7.25\\
$g$&  0.82&  7.79&  7.31&  0.72&  1.77\\
$r$&  0.93&  7.81&  7.26&  0.81&  1.69\\
$i$&  0.92&  6.84&  6.38&  0.75&  1.32\\
$z$&  0.97&  8.06&  7.61&  0.68&  2.70\\

\hline
\end{tabular}
\end{center}
\tablecomments{A summary of the calibration errors for the five SDSS filters,
as determined by simulations;
all values are in mmag. $\langle \Delta m \rangle$ is the mean of the difference between
the estimated and true calibration value for each SDSS field, while $\sigma$ is the corresponding
standard deviation, with $\sigma_{3}$ the $3\sigma$ clipped value, and $\%(3\sigma)$
the fraction (in percent) of $3\sigma$ outliers.
Finally, $\sigma_{0}$ is the 
calibration error just from measurement noise (i.e. for a simulation with no 
unmodeled random component to the atmosphere).}
\end{table}

The results of this comparison are in Table~\ref{tab:caliberr}. We start by noting 
that the calibrations are determined correctly (on average) to $\sim 0.1\%$ or better, verifying
both the algorithm and our implementation of it. The errors in
the calibrations are $< 1\%$, or $10$ mmags for all the filters (except $u$, where they are
slightly higher), suggesting that the SDSS can break the ``sound barrier'' of delivering $1\%$
relative calibrations over the entire survey region. Catastrophic failures in the calibrations
are also negligible, evidenced both from the near equality between the 
sigma-clipped and total variances, and the almost Gaussian fraction of $3\sigma$ outliers.
Finally, we note that the errors in the calibrations are dominated by the unmodeled random
fluctuations in the k-terms. Simulations with no random fluctuations achieve calibration errors
of $\sim 0.1\%$, suggesting that the SDSS calibration errors are therefore completely
dominated by unmodeled behaviour in the k-terms. The exception again is the $u$ band
where measurement noise is only a factor of $\sim 2$ smaller than the random noise 
in the atmosphere.

\begin{figure*}
\begin{center}
\leavevmode
\includegraphics[width=6.0in]{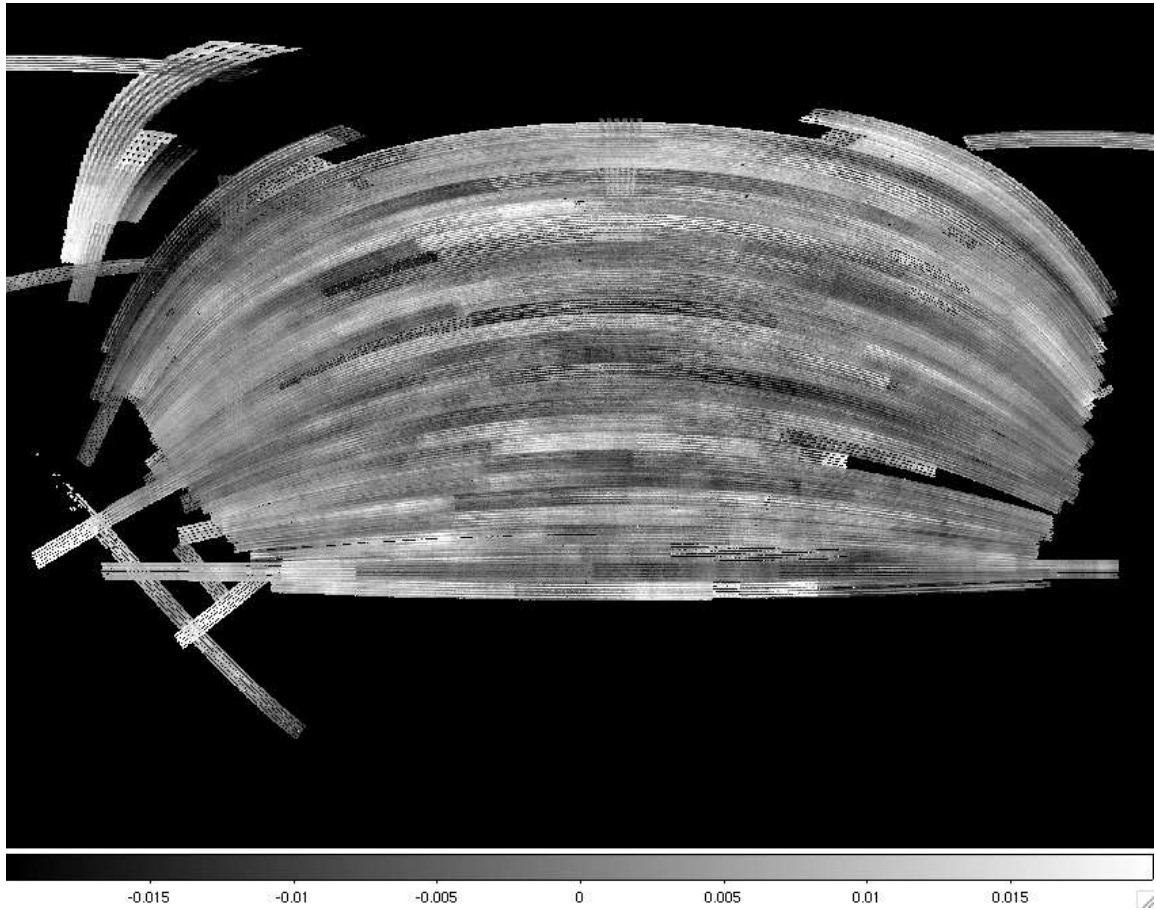}
\end{center}
\caption{An image of the calibration errors for $r$ band on the sky obtained for one
of our simulations. The projection is the same as Fig.~\ref{fig:sky_coverage}, but 
zoomed in on the Northern Galactic Cap of the SDSS. The greyscale
saturates at magnitude errors of $\pm 0.02$ mag.
}
\label{fig:s00_default0_resid}
\end{figure*}

The spatial distribution of the calibration errors is in Fig.~\ref{fig:s00_default0_resid}.
The calibrations are uniform across the whole survey area at the $\sim 1\%$ level, and 
are noticeably better at the survey poles where the number of overlap regions increases 
(see Fig.~\ref{fig:sky_coverage}). Importantly, although there is spatial structure over
individual SDSS runs (which is inevitable, given that we calibrate entire runs as atomic units),
there are no coherent structures over the entire survey region.

\begin{figure}
\begin{center}
\leavevmode
\includegraphics[width=3.0in]{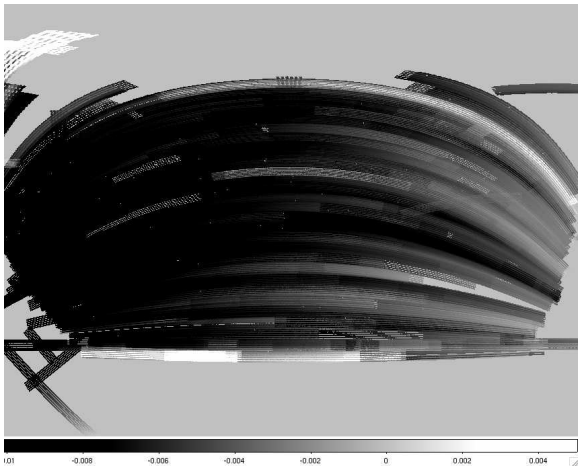}
\end{center}
\caption{The 
difference in calibration between assuming $dk/dt=0$ and the true value. 
The tilt over the survey region is clearly apparent, and is approximately
10 mmag over the survey region. The greyscale goes from $-0.01$ mag to 
$+0.005$ mag.
}
\label{fig:kdot}
\end{figure}

The above discussion assumes calibrations making the default choices 
described in Sec.~\ref{sub:implement}. We can use our simulations to discuss the 
robustness of the algorithm to these choices below. For simplicity,
we only consider the $r$ band for these tests.
\begin{itemize}
    \item \textit{Magnitude Limits :} As discussed above, the errors
    in the calibration are dominated by unmodeled systematics in the 
    atmosphere, and not measurement noise. We therefore expect the algorithm
    to be relatively insensitive to the choice of the magnitude limit. We 
    explicitly verify this by re-calibrating after
    decreasing the magnitude limit by 0.5 mag. Although this reduces the number
    of stars and observations by 30\%, the calibration errors are unaffected, as
    expected. 
    
    \item \textit{Apache Wheel data :} As described in Sec.~\ref{sec:SDSS}, the 
    SDSS imaging data was supplemented by a grid of 4x4 binned data designed to improve
    the uniformity of the calibration over the entire survey region. Calibrating the 
    survey without these data increases the calibration error to 10.4 mmag (compared with
    the 7.8 mmag in Table~\ref{tab:caliberr}), an increase of $\sim 30\%$. Most of this
    increase is however driven by catastrophic failures; the $3 \sigma$ clipped variance
    only increases to 8.1 mmag, a more modest increase of 10\%. As expected, the Apache 
    Wheel data better constrain parts of the survey that were poorly connected, 
    as they were designed to do. However,
    for regions already well constrained, the improvements are marginal.
    
    \item \textit{$dk/dt$ :} Since we do not fit for a value of $dk/dt$, we must understand
    how errors in our assumed value of $dk/dt$ propagate to the calibration. Fig.~\ref{fig:kdot}
    shows the difference between calibrating a simulation assuming the correct value of 
    $dk/dt$, and assuming $dk/dt=0$. While the increase in the size of the calibration errors
    is small, the incorrect value of $dk/dt$ introduces an overall tilt to the survey
    (in the figure, this is approximately 10 mmag). This tilt results from 
    the fact that regions of similar RA are observed at approximately the same relative
    time in the night. The errors from an incorrect $dk/dt$ therefore do not cancel, 
    but accumulate into a tilt, because we always observe the sky west to east.
    This is exacerbated by the fact that there is little data connecting the survey at the ends
    through the Galactic plane, and therefore no closed loops to prevent the appearance of such 
    a tilt.
    This is the most serious systematic error in the 
    calibration, and could affect any large scale statistical measures. In fact, both
    \cite{2007MNRAS.378..852P} and \cite{2007MNRAS.374.1527B}
    observe excess clustering of photometrically 
    selected luminous red galaxies at the very largest scales. We speculate that a tilt
    in the calibration could be a possible contaminant to the measurements on those 
    scales.

\end{itemize}

\section{The SDSS Photometric Calibration}
\label{sec:sdssresults}

Having described and verified our algorithm, we apply it to the 
SDSS imaging data. Since we do not have ground truth to compare our results, 
we describe both the internal consistency (Sec.~\ref{sub:internal}) 
and astrophysical tests (Sec.~\ref{sub:principal_colors}) we use to assess the 
photometric calibration. In addition, we also address the spatial structure
of the calibration errors (Sec.~\ref{sub:errormodes}), as well as the photometric
stability of the SDSS (Sec.~\ref{sub:stability}). Finally, we compare our calibrations
with the currently public SDSS calibrations (Sec.~\ref{sub:comparison}).

In what follows, we use ``magnitude residual'' to denote the difference between 
the (calibrated) magnitude of an observation of a star and the mean magnitude of 
all observations of the star.

\subsection{Internal Consistency}
\label{sub:internal}

\begin{figure}
\begin{center}
\leavevmode
\includegraphics[width=3.0in]{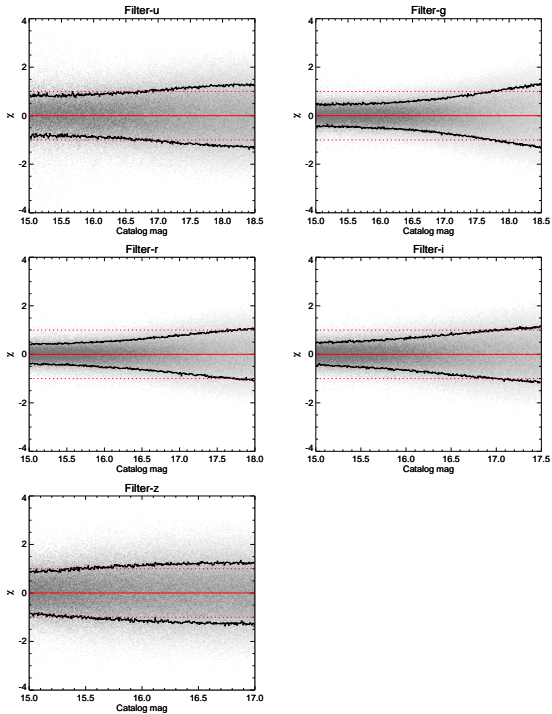}
\end{center}
\caption{The magnitude residuals weighted by their errors ($\chi$)
as a function of apparent magnitude, for the 
five SDSS filters. The residual for a given observation is the difference between 
the observed magnitude and the mean magnitude averaged over all the observations
of the star. The dotted lines show $\chi = \pm 1$, while the solid lines show the 
16\% and 84\% contours; these should coincide with 
the $\chi= \pm 1$ lines if the scatter in the magnitudes is well described by the
errors. The discrepancy at bright magnitudes is due to an error floor we impose
to down-weight the brightest stars. 
}
\label{fig:mresid}
\end{figure}

The first internal consistency test is the distribution of magnitude residuals. 
Since the scatter in the residuals also includes measurement noise ($\sigma$), 
it is more illuminating to consider $\chi = (m-\langle m \rangle)/\sigma$; if the measurement
errors are a good estimate of the scatter in the residuals, $\chi$ should be Gaussian distributed 
with unit standard deviation. This is plotted for the stars used in the calibration,
for the five filters, in Fig.~\ref{fig:mresid}. At the faint end, we observe that $\chi$
is distributed as expected, suggesting that the measurement noise is a good description of the 
scatter, and that calibration errors do not appreciably increase the scatter. The discrepancy at 
the bright end is due to a floor ($\sigma = 0.01 {\rm mag}$ added in quadrature) 
we impose on the magnitude residuals, to reflect the fact that
the dominant error for these stars is no longer Poisson noise but possible systematics in the
measurements. Note that calibration errors would only broaden the distribution of $\chi$.

\begin{figure}
\begin{center}
\leavevmode
\includegraphics[width=3.0in]{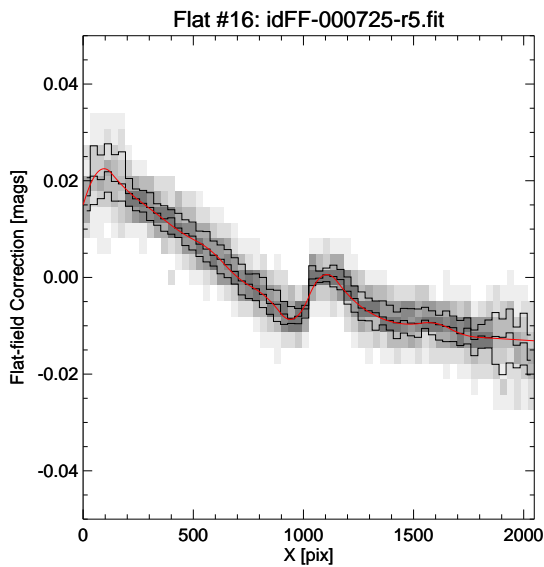}
\end{center}
\caption{An example of a flatfield vector from one $r-$band chip during
Season 3. The grey scale and 25\%, 50\%, and 75\%
contours show the magnitude residuals as a function of CCD column, for all stars 
observed multiple times during that season. The smooth central (red) line shows the 
best fit (splined) flatfield vector.
}
\label{fig:plotflat}
\end{figure}

We also consider the magnitude residuals as a function of the CCD column, grouping
the data by CCD and flat field seasons; this is an estimate of the accuracy of our
flat field correction.
An example of these magnitude residuals as a function of the
CCD column for camcol 5 in $r-$band is shown in Fig.~\ref{fig:plotflat}. 
We don't correct for the flat 
field in this plot, to show the structure of the flat field itself.
The r.m.s scatter in the magnitude 
residuals about the derived vector is $\sim 0.5$\% throughout the chip, although
it increases at the edges of the CCD. Also, since we do not fit for amp-jumps
but use the values derived from the photometric pipeline, the flat fields adjust to 
correct for errors in the amp-jumps. Note that the errors in the amp-jump estimation 
are small (the true amp-jumps are usually a few tenths of a magnitude, while the 
errors are a few millimagnitudes) that the splines have the necessary
flexibility to adequately flatten the field.

\begin{figure}
\begin{center}
\leavevmode
\includegraphics[width=3.0in]{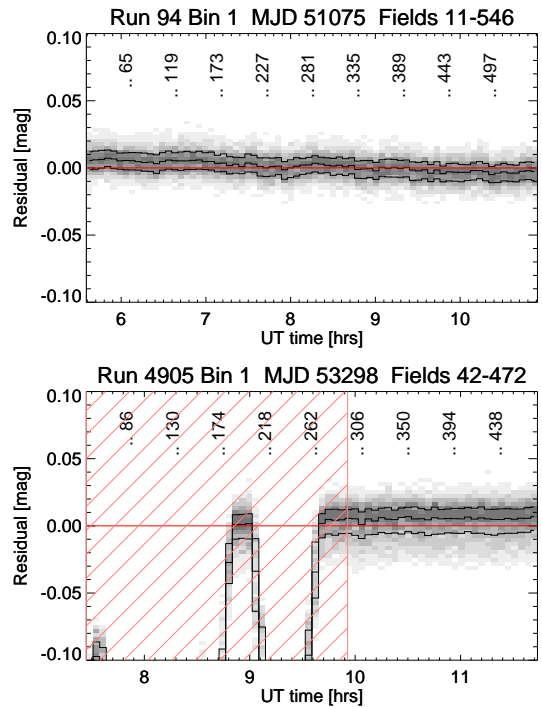}
\end{center}
\caption{Magnitude residuals as a function of time/field number for two example 
runs; all six camera columns are combined in these plots. The contours again show 
the 25\%, 50\%, and 75\% levels. The red hatched regions mark periods of time 
independently known to be non-photometric from the SDSS photometricity monitors.
Note that both these runs are on the multiply imaged Equatorial stripe, and therefore
have lots of overlaps. However, for a large fraction of the data, the overlaps
are considerably more sparse.
}
\label{fig:plotrunresids}
\end{figure}

Finally, we plot the magnitude residuals, grouped by run,
as a function of field number (and time); two examples are in Fig.~\ref{fig:plotrunresids}.
These plots are our primary diagnostic of the photometricity of the data. Photometric
data have the mean residual scattered around zero, although often with coherent errors at
the few millimagnitude level. By contrast, the residuals for unphotometric data show large 
excursions from zero, often at the $\sim 10\%$ level or greater. Most of these data have
already been correctly flagged as being non-photometric by the SDSS photometricity monitors 
\citep{2001AJ....122.2129H} and have been excluded from the solution. Any remaining 
non-photometric data is manually flagged as such, and removed in a second iteration of the
calibration. For all the non-photometric data that overlaps photometric data, we can estimate
an a-term per field that minimizes the residuals which determines the calibration of those
fields (these are still flagged as being non-photometric).

\subsection{Spatial Error Modes}
\label{sub:errormodes}

\begin{figure}
\begin{center}
\leavevmode
\includegraphics[width=3.0in]{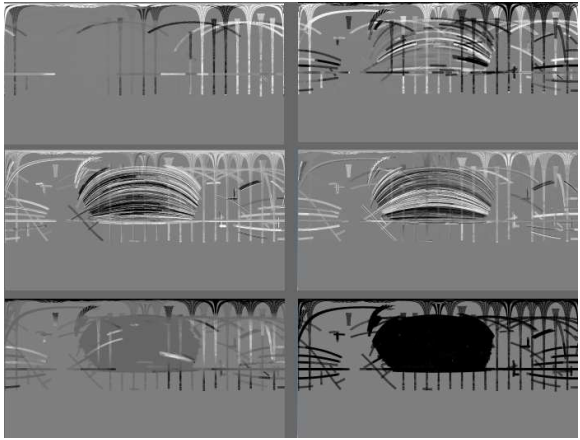}
\end{center}
\caption{Examples of the spatial structure in the calibration errors for the 
$r$ band, organized from
left to right, and top to bottom in increasing order of their uncertainties. The top 
left mode is the best constrained, while the bottom right mode is the worst constrained.
The middle row are examples of modes with typical errors. The modes are normalized such 
that the maximum absolute error is 1. Note that the worst constrained mode is the 
exactly degenerate overall zeropoint of the survey. The structures are similar for the
other bands.
}
\label{fig:modes}
\end{figure}

Since our goal in this paper is accurate relative calibration, it behooves us to 
understand the spatial structure in the calibration errors. Our starting point
is the curvature matrix, Eq.~\ref{eq:chi2_curve}. The eigenvectors of this matrix
partition our basis of photometric parameters into uncorrelated linear combinations,
whose uncertainties are given by the inverses of the corresponding eigenvectors.
An error in each photometric parameter can be thought of as a pattern of errors 
on the sky, determined by the runs corresponding to that parameter. One can use this
to project the eigenvectors (modes) of the curvature matrix on the sky.
These then describe the spatial structure 
of the calibration errors (Fig.~\ref{fig:modes}).
Note that projecting these modes on the sky destroys the
linear independence of the modes; if desired, this can be restored by a straightforward
orthogonalization. 

The worst constrained mode is, as expected, the zero point of the calibration which is 
exactly degenerate. However, 
examining the other poorly constrained modes (an example of which is the bottom left of 
Fig.~\ref{fig:modes}), we observe that there are no other such simple large scale modes,
an indication of the fact the survey is well connected.
At the other extreme are the best constrained modes. These are typically complicated combinations,
and not surprisingly describe modes held together by the grid of Apache Wheel data.
More illuminating are examples of typical modes, two of which are in the 
middle row of Fig.~\ref{fig:modes}. The most noticeable characteristic is the striping
along the scan direction. This simply reflects the fact that we calibrate camera columns
individually, resulting in errors correlated in the scan direction.

We do not fit for $dk/dt$, and so it does not 
get included in the curvature matrix. However, we saw in Sec.~\ref{sub:simresults} 
that it resulted in a coherent tilt from one end of the survey to the other.

\subsection{Experimental Stability}
\label{sub:stability}

\begin{figure}
\begin{center}
\leavevmode
\includegraphics[width=3.0in]{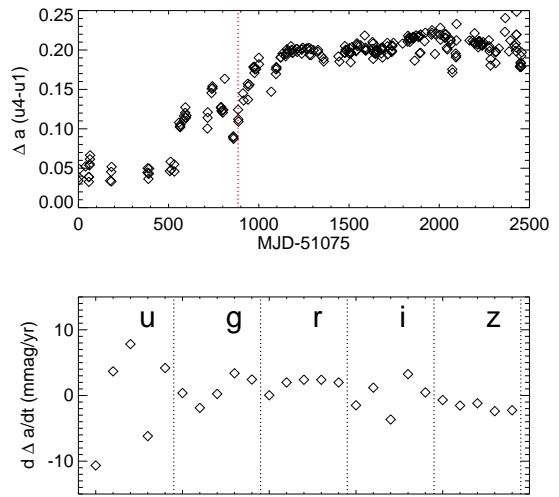}
\end{center}
\caption{[top] The difference between the a-terms of $u$ band camera
column 4 and 1; since the atmospheric corrections are common to both of these, 
this is a measure of the stability of the telescope + camera system. The 
drift in the camera during early data due to problems with the vacuum system
(see Table~\ref{tab:flatseasons}) is clearly visible; the vertical dotted line
at MJD 51960 marks when the vacuum system was fixed. Note that all of the changes are
long term drifts; the system is stable on short (i.e. day) intervals as 
assumed in our model. [bottom] The drift in the relative (to camera column
1) zero points of camera columns 2 through 6, for all five filters in 
mmag/year, measured after MJD 51960.
}
\label{fig:camera_stable}
\end{figure}

Our results for the photometric calibration can also be used to estimate the 
overall photometric stability of the SDSS camera, telescope and site. We estimate
the camera stability by considering differences between a-terms as a function of 
time (for definiteness, we compute the a-terms relative to camera column 1); these 
differences are insensitive to any common mode effects (such as the atmosphere). An
example is in Fig.~\ref{fig:camera_stable}. During the initial phases of the survey, 
we note that the camera was not very stable over long time periods, reflecting 
various problems with the vacuum system flagged in Table~\ref{tab:flatseasons}.
However, over the past $\sim 5$ years, the camera has been extremely stable, as
evidenced by an overall drift in the a-term differences of $< \sim 10$ mmag/yr, for 
all the CCDs. 

\begin{figure}
\begin{center}
\leavevmode
\includegraphics[width=3.0in]{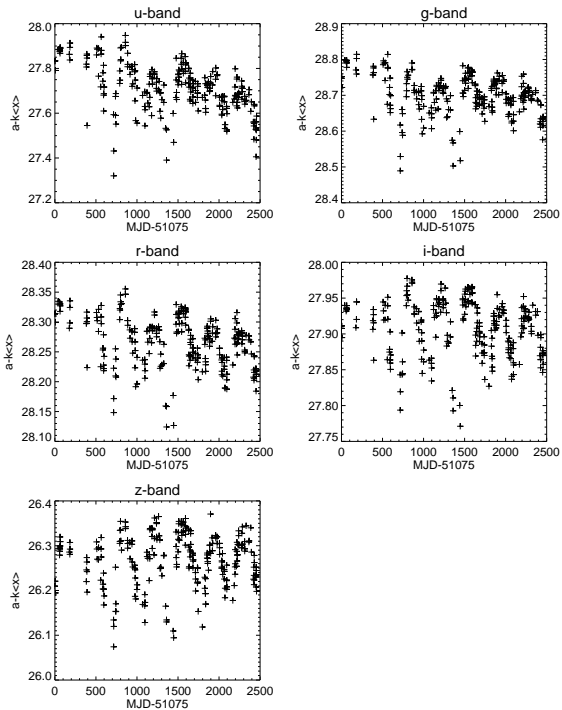}
\end{center}
\caption{The value of $a- k\langle x \rangle$ for camera column 1 
as a function of MJD, and filter; $\langle x \rangle$
is the mean airmass of all the observations in a given night. This combination 
is insensitive to degeneracies between the a- and k-terms, and measures the 
overall photometric stability of the SDSS camera, telescope and site. The
seasonal variations in these data are clearly apparent, as is the fact that 
the mirror is aluminized every summer.
}
\label{fig:site_stable}
\end{figure}

One could also measure the combined stability, treating the camera, telescope, and site
as a combined system. As the a- and k-terms are degenerate, we consider the combination
$a-k\langle x \rangle$ every night, where we average the airmass over all the observations
in a given night. This is plotted in Fig.~\ref{fig:site_stable} for the five SDSS filters.
The most striking aspect of these data are the seasonal variations, seen as periodic 
oscillations in the data, at the $\sim 10\%$ level (except in the $u$ band, where they are
$\sim 20\%$). Factoring out the seasonal variations, we find less than a 5\% drift over the 
$\sim 7$ years considered here (again, except the $u$ band, where the drift is $\sim 10\%$
over the same time period).

We emphasize that all of the effects discussed here are long-term
effects and do not affect the quality of the calibration which only assumes
stability over a night for the a- and k-terms.

\subsection{Principal colors}
\label{sub:principal_colors}

The above discussion has relied on a combination of simulations and internal consistency checks 
to assess the quality of the calibrations. While these provide essential perspectives, they have 
important disadvantages as well. 
Internal consistency checks are not independent of the calibration and might not flag
deviations from the input model. Furthermore, these checks are local measurements, and do 
not provide information about large scale systematics problems. While simulations fill that
gap, they are limited by the input model used. Astrophysical tests complement the above by 
providing large scale, independent verification, and are ultimately limited by astrophysical uncertainties.

The majority ($> \sim 98\%$) of the stars detected by the SDSS are on the main sequence
\citep{2000AJ....120.2615F,2003ApJ...586..195H}, and lie on one-dimensional manifolds (the ``stellar locus'')
on color-color diagrams. This suggests using the position of the stellar locus as a diagnostic
of calibration errors \citep{2004AN....325..583I}. While there are a number of morphological
features one could use as a marker, we follow the discussion in \cite{2004AN....325..583I} 
and use 
the ``principal colors'' that define directions perpendicular to the stellar locus. We 
consider four such colors \citep{2004AN....325..583I} : $s$ (perpendicular to the blue
part of the locus in $u-g$ vs. $g-r$ plane), $w$ (the blue part in $g-r$ vs. $r-i$), 
$x$ (the red part in $g-r$ vs. $r-i$), and $y$ (the red part in $r-i$ vs. $i-z$):
\begin{eqnarray}
\label{eq:princip_color_def}
s & = & -0.249 u  + 0.794g -0.555 r + 0.234 \nonumber \\
w & = & -0.227 g + 0.792r - 0.567i + 0.050 \nonumber \\
x & = & 0.707 g  - 0.707 r - 0.988 \nonumber \\
y & = & -0.270 r + 0.800 i - 0.534 z + 0.054 \,\,. 
\end{eqnarray}
We correct all magnitudes with the \cite{1998ApJ...500..525S} estimates
of extinction (except immediately below), but do not attempt any
correction for stars not completely behind all the dust.
Since we calibrate each band separately, and apply no color cuts
to select the stars used, the above principal color diagnostics provide 
a completely independent verification of the calibration. 

\begin{figure*}
\begin{center}
\leavevmode
\includegraphics[width=6.0in]{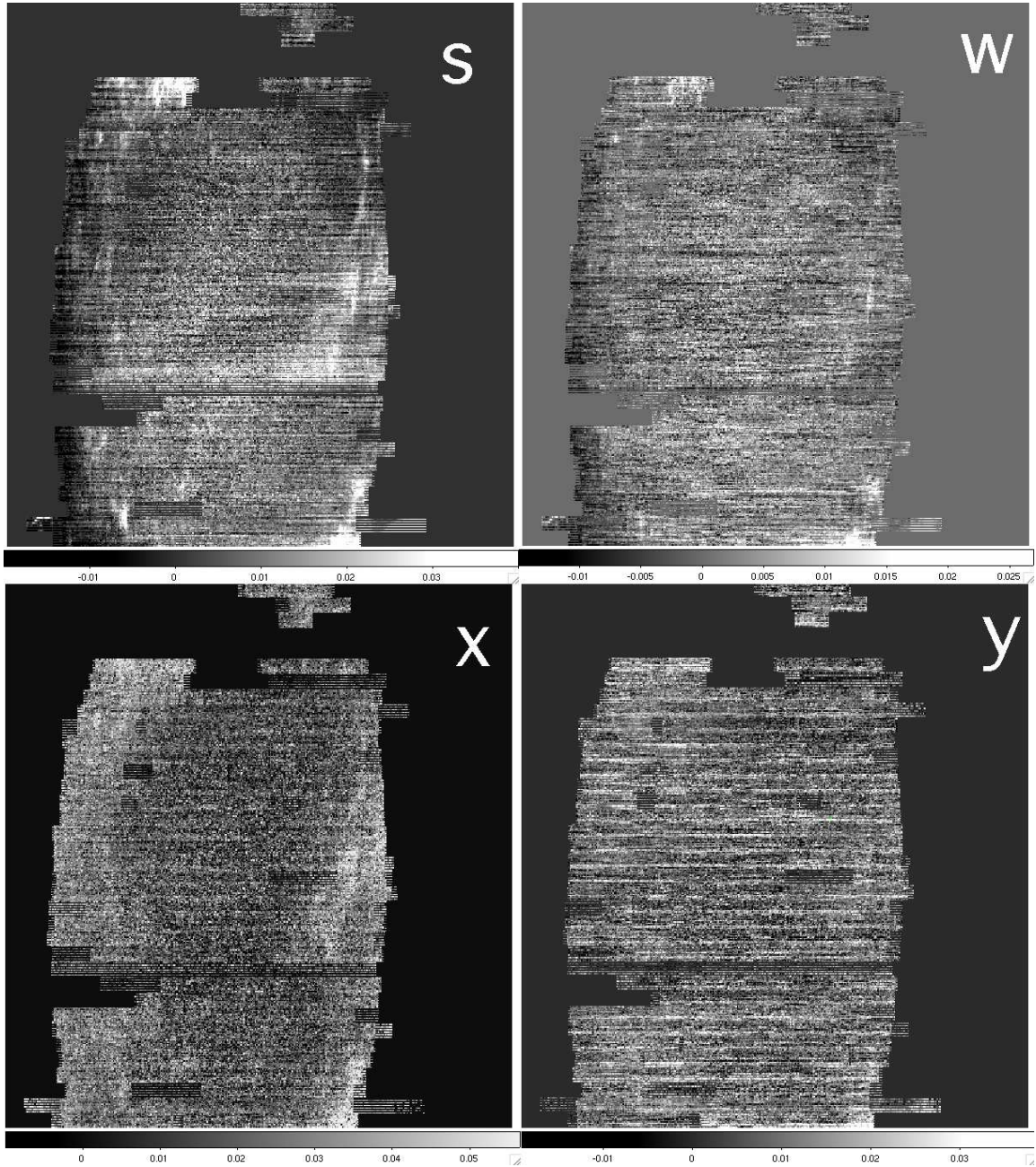}
\end{center}
\caption{The spatial variation in the $s$ (top left), $w$ (top right), $x$ (bottom left),
and $y$ (bottom right) principal stellar colors. The projection is a $\mu-$stripe projection,
with the $x$ coordinate measuring $\mu$ (the coordinate along the scan direction). The $y$ coordinate
is the SDSS stripes, with each row as one of the 12 camera columns that define the stripe.
We have restricted ourselves to data on survey stripes between 9 and 44, corresponding to most of the 
North Galactic cap, in this plot. Note that the aspect ratio in the
$\mu$ direction is significantly compressed.
}
\label{fig:pcolor2d}
\end{figure*}

Fig.~\ref{fig:pcolor2d} plots these on a $\mu$-stripe projection; the $x$-direction is the coordinate
along the scan direction $\mu$ \citep{2003AJ....125.1559P}, while the $y$ coordinate is given by 
\bea
y & = 12({\rm stripe}) + 2({\rm camcol})-2 \,, &{\rm strip=S} \nonumber \\
& = 12({\rm stripe}) + 2({\rm camcol})-1 \,, & {\rm strip=N} \,\,,
\eea   
where stripe, strip and camcol is the SDSS stripe number, whether it is a northern
or southern strip, and the camera column respectively. This lays out
each camera column as a row, respecting the interleaved structure of the strips within
a stripe. The advantage of this projection is that calibration errors appear principally as 
stripes in the $\mu$ direction,  while Galactic structure appears as irregular
structures localized in $\mu$, making it easier to separate the two. 
For the purpose of this plot, we use colors not corrected for extinction
to highlight the Galactic structure.
We simply exclude the 
small fraction of data not on survey stripes for the purposes of this analysis.

\begin{figure}
\begin{center}
\leavevmode
\includegraphics[width=3.0in]{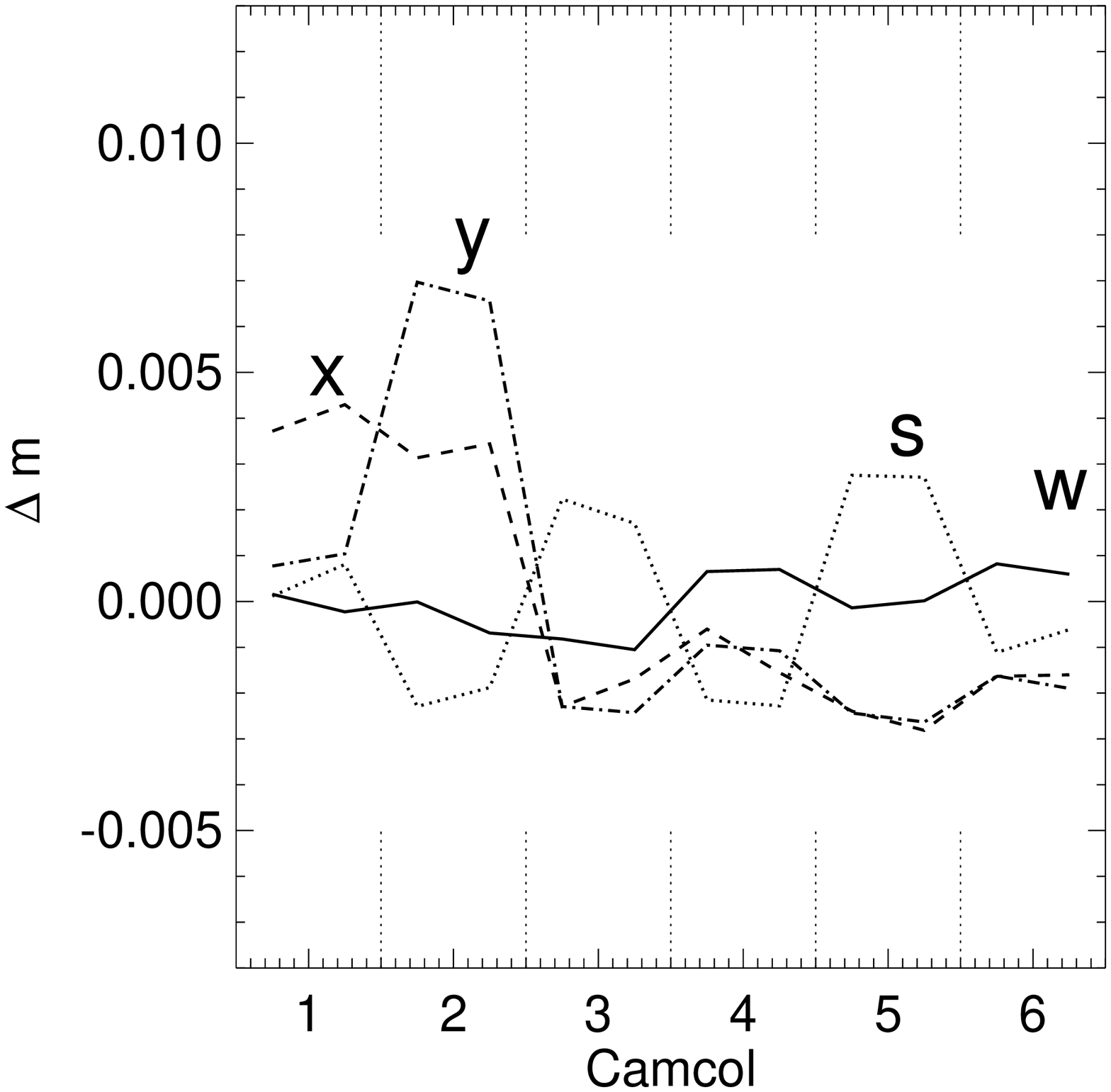}
\end{center}
\caption{The average principal colors measured for each of the 6 camera columns, 
for the north and south strips separately. As before, we restrict ourselves to stripes between
9 and 44. The mean color has been subtracted from 
each of the four curves; the means are -0.002, 0.004, 0.007 and 0.007 for $s$, $w$, $x$ and
$y$ respectively. The variations between camera columns are  $<1\%$ (peak to peak) for all colors.
}
\label{fig:pcolor_striping}
\end{figure}

We note that there is little visual evidence for any striping over the entire survey region
in $s$, $w$, and $x$. Reddening from Galactic dust is clearly visible in $s$ and $x$, which
are colors nearly parallel to the reddening vector. The $y$
map, on the other hand, does appear to show striping, with a periodicity on the SDSS stripe scale.
In order to quantify this effect, we plot the average principal color per camera column (distinguishing
between northern and southern strips) in Fig.~\ref{fig:pcolor_striping}.
As anticipated from the 2D maps, the $s$, $w$, and $x$ colors are uniform at the $0.5\%$ (peak to peak) level, 
whereas camera column 2 is offset in $y$ at 0.7\% (peak to peak). It is unlikely that this is an 
artifact of the calibration process which treats all camera columns identically. Since $y$ is 
the only color to use the $z$ band, we speculate that this could be caused by the known variations 
in the $z$-band filter responses.
However, as this effect is of the same order as other systematics
present in the calibration (and below our target of 1\%), we simply caution the reader about 
this systematic in this paper, and defer its resolution to future work.

\begin{figure}
\begin{center}
\leavevmode
\includegraphics[width=3.0in]{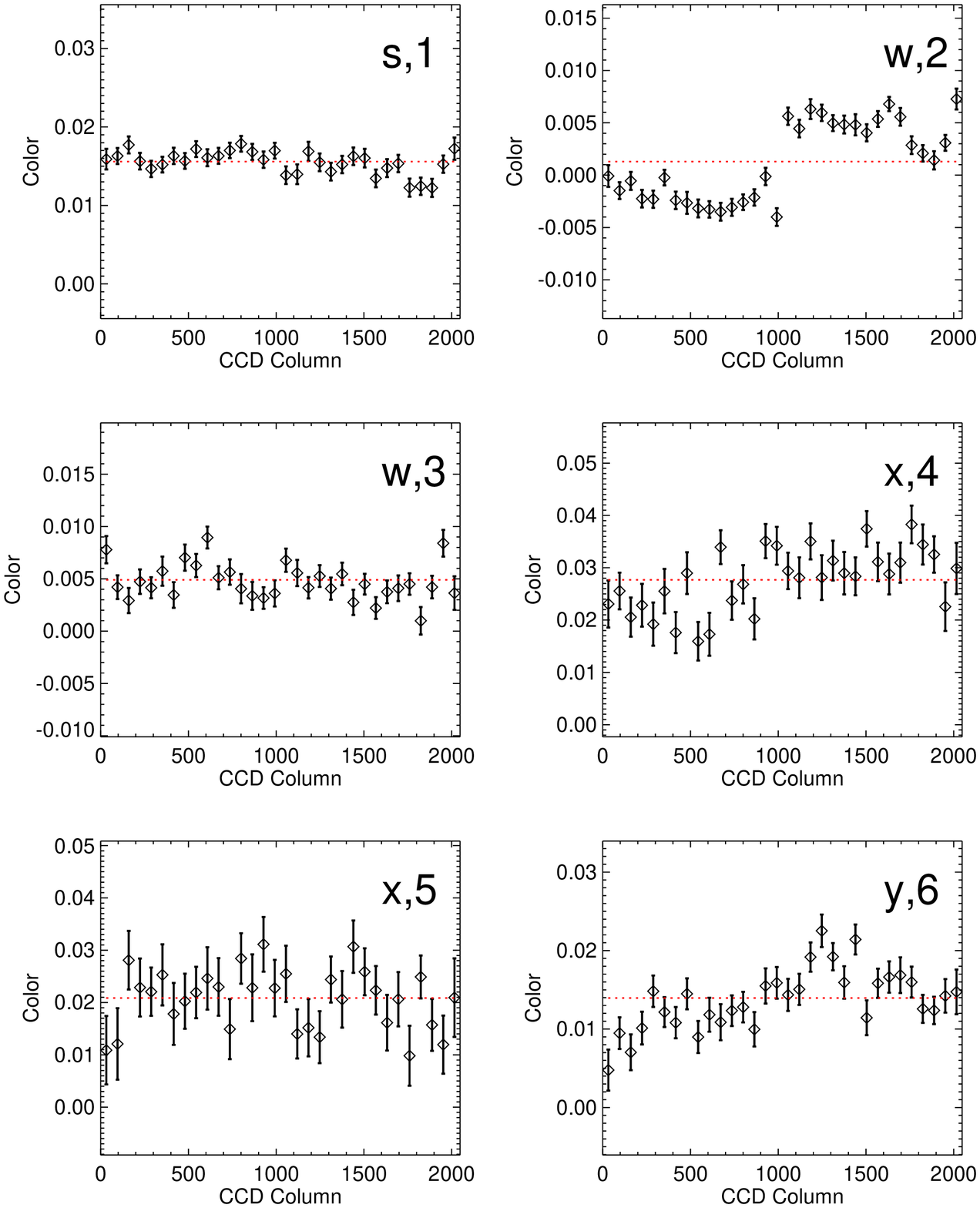}
\end{center}
\caption{Examples of the principal colors as a function of CCD column.
The principal color and camera column used is noted in each panel. Each panel 
plots a color measured over a group of runs, chosen to be in the 
same flatfield season (Table~\ref{tab:flatseasons}). From 1 through 6, the runs
used are (745, 752,756),(1331,1345),(2190, 2299),(2566, 2662, 2883, 2886), (3560, 3830)
and (4927, 5052) respectively. Deviations in the color from a constant (dotted line) indicate
errors in the flatfield determination.
}
\label{fig:pcolor_flat}
\end{figure}

The above has focused on large scale systematics; Fig.~\ref{fig:pcolor_flat} shows examples
of the principal colors as a function of CCD column averaging over a random sample of 
runs in a flat field season. The deviations from a constant color are $\approxlt 1\%$ for all colors,
and $<0.5\%$ for $s$ and $w$, consistent with our estimates from simulations. We also observe
errors in the amp-jump determination at the $\sim 0.5\%$ level, similar to Fig.~\ref{fig:plotflat}.

\subsection{Comparison with previous results}
\label{sub:comparison}

\begin{figure}
\begin{center}
\leavevmode
\includegraphics[width=3.0in]{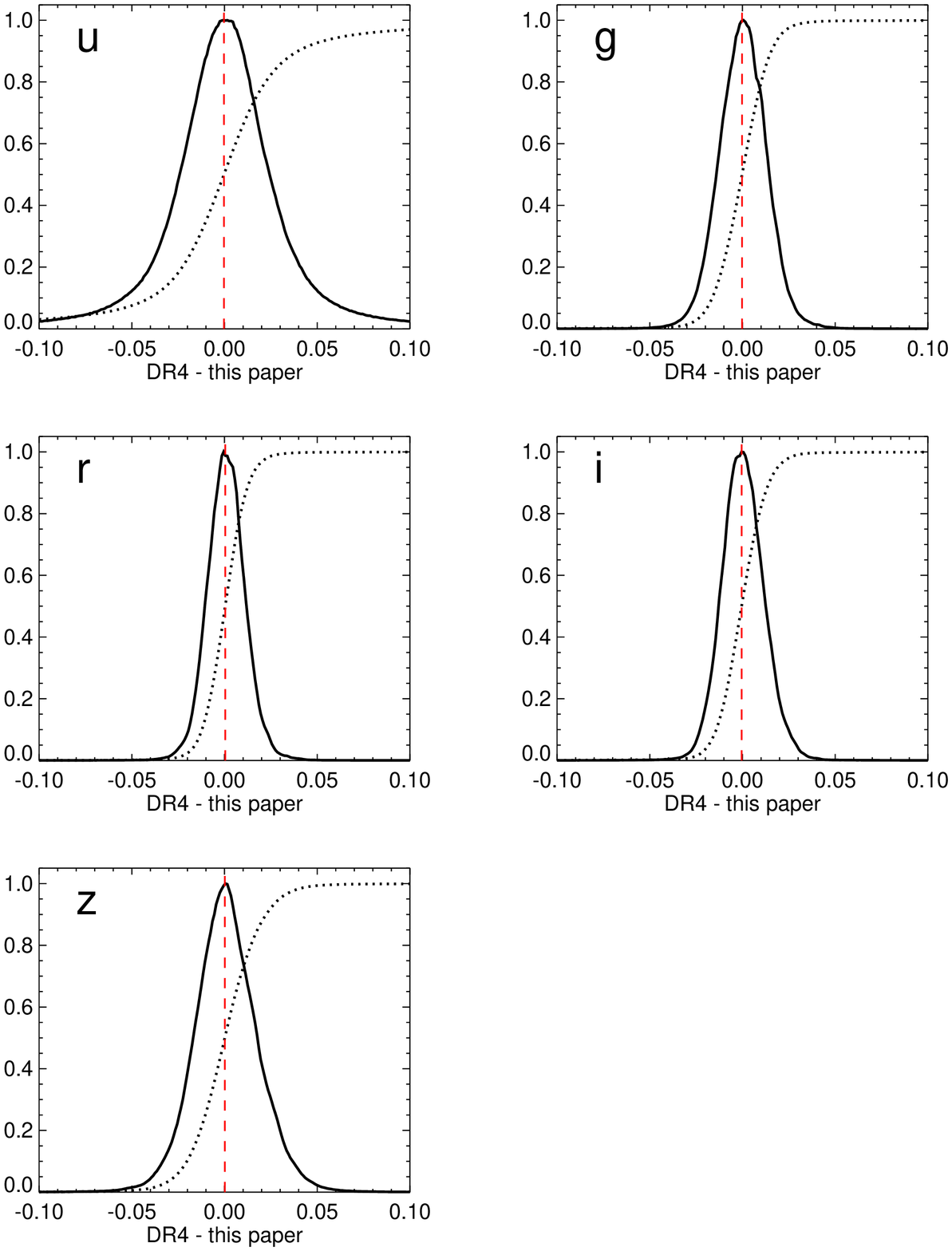}
\end{center}
\caption{Histograms of the differences in magnitudes between DR4 and this work,
for stars with $r$-band magnitudes $<$ 18.0. Shown are the distribution of differences
(normalized to have a maximum value of 1),
as well as the cumulative distibution (dotted line). The vertical lines
show the median of the distribution, which is $< 0.001$ mag for all five filters.
}
\label{fig:fnal_hist}
\end{figure}

We conclude this section by comparing the calibrations presented here with those publicly
available as part of Data Release 4 \citep[DR4]{2006ApJS..162...38A}
\footnote{\texttt{http://www.sdss.org/DR4}}. 
Fig.~\ref{fig:fnal_hist} shows the difference between
the aperture magnitudes of DR4 and those derived in this paper, for all stars with $r$ band
magnitude less than 18. The magnitudes agree on average 
by construction, as the zero points were determined 
by matching to the public calibration. Furthermore, the scatter is approximately 2\% (r.m.s)
for $griz$, and 3\% (r.m.s) in $u$, consistent with the published uncertainties. The Data Release 
errors are therefore dominated by the Photometric Telescope (PT) based calibration method.

\begin{figure}
\begin{center}
\leavevmode
\includegraphics[width=3.0in]{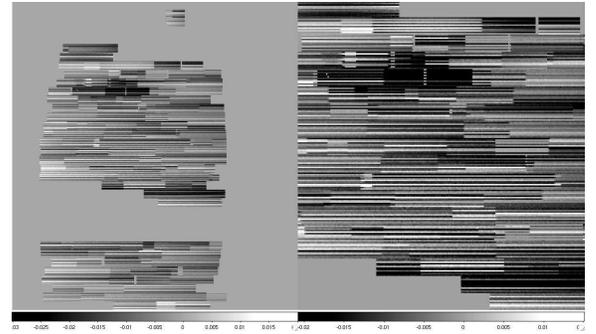}
\end{center}
\caption{The difference in magnitudes  between DR4 and this work in the $r$ band for 
the stars in Fig.~\ref{fig:fnal_hist},
plotted in the $\mu$-stripe projection. The right panel zooms in to a region to highlight the
structure in the calibration errors.
}
\label{fig:fnal_compare2d}
\end{figure}

Fig.~\ref{fig:fnal_compare2d} plots these differences in the $\mu$-stripe projection introduced
previously. Since the standard SDSS calibration does not attempt to explicitly control relative
calibration errors, the striping in the figure is not surprising. Note that the errors are correlated
in the $\mu$ direction as expected, but also across camera columns. The latter arises from the fact 
that the calibration patches are 40' wide and span three camera columns, thereby correlating their calibrations.

\begin{figure}
\begin{center}
\leavevmode
\includegraphics[width=3.0in]{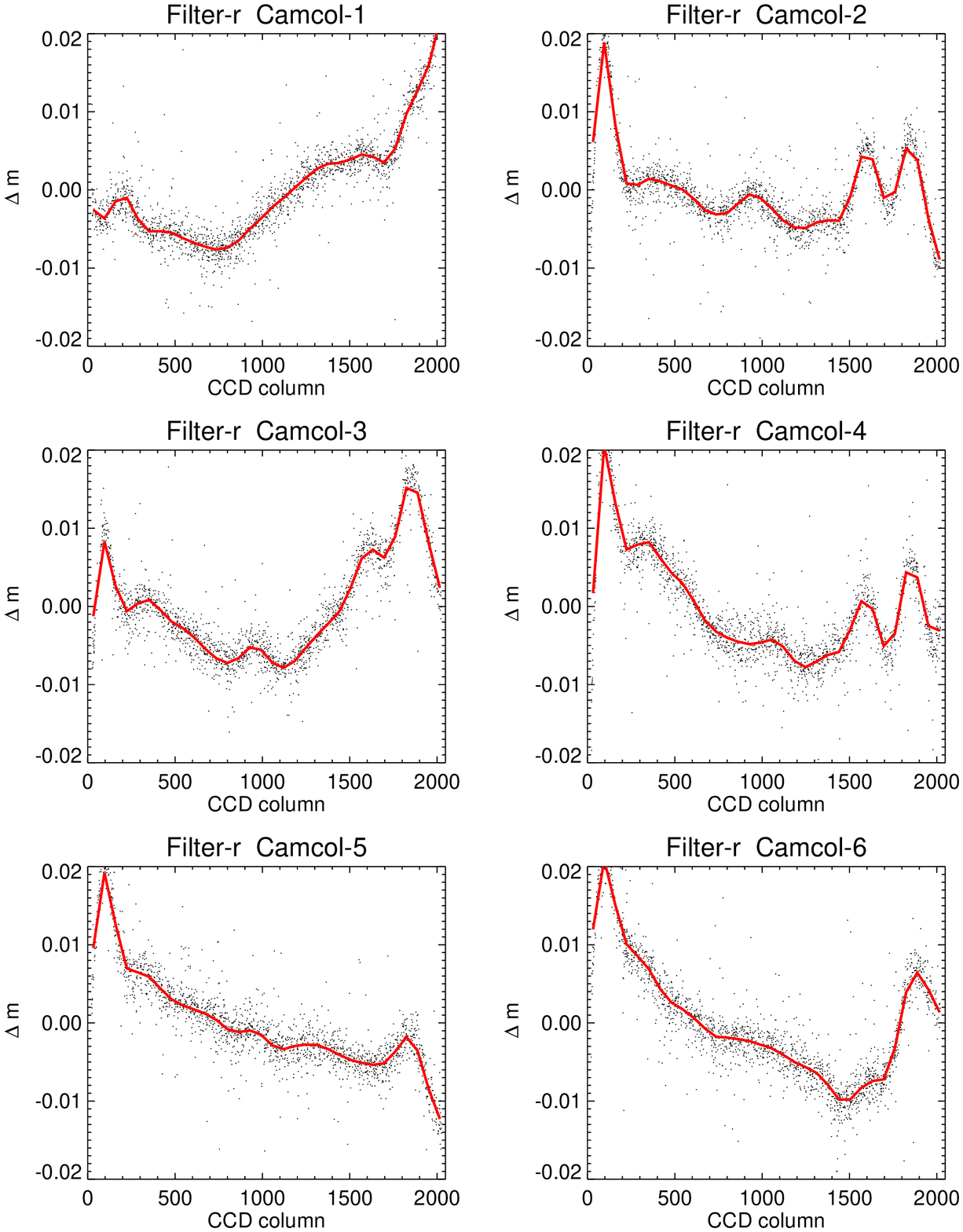}
\end{center}
\caption{The difference between the $r$ band DR4 flat fields and those determined in this
work, for SDSS runs between 4100 and 4400. These differences are traced by 
the difference in magnitudes as a function of CCD column,
for the 6 camera columns. The solid line shows the median difference, while the points
are a subsampling of the individual measurements. We restrict ourselves to runs between
4100 and 4400 to select runs within a single flat field season.
}
\label{fig:fnal_flat}
\end{figure}

Finally, Fig.~\ref{fig:fnal_flat} plots the differences in the DR4 flat fields, and those 
determined in this paper, for an example flat field season. The errors in the flat fields
are both higher than the quoted uncertainties and appear to have long wavelength power.
We speculate that these result from the method used to determine the flux response of the 
CCDs, which aliases flat field errors in the Photometric Telescope into the 
final flat fields. This aliasing is mitigated by using the average of $g$,$r$, and $i$,
instead of any of those bands individually; this does not however eliminate the problem.
Formally, these errors are $\sim 1\%$ (r.m.s.), but are highly correlated, both spatially and in 
color.
 
\section{Public Data Release}
\label{sec:public}

The calibrations (dubbed ``ubercalibration'') 
described in this work have been made public with the SDSS
Data Release 6 \footnote{\texttt{http://www.sdss.org/dr6}}, and will be updated for
the subsequent data releases. The SDSS Catalog Archive Server has recalibrated 
versions of the most popular magnitudes, as well as correction terms that
can be applied to other magnitudes. We refer the reader to the documentation
under ``ubercalibration'' on the SDSS data release websites for the most
up-to-date information on these calibrations, and the available data formats.

\section{Discussion}
\label{sec:discussion}

We have presented a method for calibrating wide field imaging surveys using overlaps in 
observations, and applied it to 
the Sloan Digital Sky Survey imaging. 
Early versions of these calibrations have already been used for the creation 
of a number of auxiliary SDSS catalogs 
\citep[e.g.][]{2004AJ....128.2577F,2005AJ....129.2562B}
as well as a number of SDSS scientific publications 
\citep[e.g.][]{2004ApJ...606..702T, 2005ApJ...633..560E, 2007MNRAS.378..852P, 2006PhRvD..74l3507T}.

The principal features and results of this work are :
\begin{itemize}
    \item {\it Relative vs. Absolute Calibrations:} We explicitly separate the problem of relative
    calibrations from that of absolute calibration. 
    The problem of absolute calibrations then reduces
    to determining one zeropoint per filter for the entire survey
    (however, see the caveat on spectral energy distributions below), 
    and does not alias into spatial variations
    in the calibration. This allows us to better control and quantify the errors in the relative
    calibration.
    
    \item {\it Simulations :} We emphasize the utility of simulations, both to validate pipelines,
    and to quantify the structure in the calibration errors. Simple analytical estimates are
    insufficient to characterize the errors, while astrophysical estimates are limited
    by their intrinsic scatter. Developing realistic simulations for the next generation 
    of surveys must be an essential part of any calibration pipeline. Such simulations
    also are invaluable in determining the observing strategy 
    ({\it before} any data are actually taken) that yields the desired
    calibration accuracy.
    
    \item {\it Stellar Flat Fields :} The problems of flat fielding wide field-of-view instruments,
    namely (i) non-uniform illumination for dome flats, (ii) spatial gradients and scattered light for sky flats
    and  (iii) mismatched spectral energy distributions (SEDs)
    have been discussed extensively in the literature 
    \citep[e.g.][]{1995A&AS..113..587M, 1996PASP..108..944C,2004PASP..116..449M, 2006ApJ...646.1436S}. 
    In particular, initial attempts to use sky flats for the SDSS resulted in errors of 
    5\% in the $r$ band and as bad as 20\% in the $u$ band, due to scattered light within
    the instrument. These were therefore never used for the public data; instead the 
    published SDSS flat fields are determined from the position of the stellar locus.
    The use of stellar flat fields mitigates all three of these \citep{1995A&AS..113..587M}. 
    Given sufficient observations and overlaps, one has sufficient S/N
    to map out the entire flat field with high precision. Furthermore, since one is using a
    realistic ensemble of stars by construction, biases due to differences 
    between the flat field SED and the SED of a given object are minimized. Note that there are 
    still potential biases for objects of unusual color; these cannot however be treated in a general manner.
    
    \item {\it 1\% Relative Calibrations/Spatial Error Modes :} Our recalibrated SDSS imaging data has
    relative calibration errors, determined from simulations, of $\sim$ 13, 8, 8, 7, and 8 mmag in $ugriz$ 
    respectively. We however do detect systematics not modeled in our simulations at the $\sim 0.5$\% level,
    suggesting a conservative estimate of $1\%$ errors in $griz$ and $2\%$ in $u$.
    In addition, we are able to characterize the spatial structure
    of the errors as a combination of error modes. Most of these modes show little coherent spatial 
    structure. The most significant spatial structure results from misestimating the 
    time variation of the k-terms, which introduces a tilt into the survey.

\end{itemize}

Throughout this paper, we ignored a number of sub-dominant systematic effects.
We briefly discuss these below, both to document
their existence as well as to alert future surveys of potential pitfalls.
\begin{itemize}
    \item {\it Spectral Energy Distributions:} When interpreting our magnitudes as
    absolute, our algorithm implicitly assumes that all objects
    have the same spectral energy distribution. The median $r-i$ color of stars used 
    in the calibration is $\sim 0.2$, and we expect the calibrations to be accurate (at the 
    stated levels) for objects with colors not very different from these stars. 
    This will however {\it not} be true for objects with unusual SEDs (e.g. SNe).
    In these cases, one must integrate the SED over the system response, in order
    to get an absolute, calibrated flux $(\rm{in erg\, cm}^{-2} \rm{s}^{-1})$.
    
    \item {\it Filter Curves:} We assume that the six copies of each filter are 
    identical, and do not attempt any color corrections between the six camcols. 
    We verified the validity of this assumption by generating synthetic $ugriz$ photometry
    for the Gunn-Stryker spectra \citep{1983ApJS...52..121G} for each of the six 
    camcols, using the individually measured filter curves. For stars with median $r-i$
    color close to $\sim 0.2$, the differences between the various camcols was better
    than $1\%$ for $griz$ and $\sim 1\%$ for the $u-$band. Of $griz$, the most drastic variation with
    color occurs for the $z-$band, with a $0.01$ mag gradient between $r-i=0$ and $1$
    seen is almost all the camcols. Gradients of a similar magnitude are also seen for $g2$,
    $g4$ and $r3$. 
    
    \item {\it Absolute Calibrations:} We ignored the issues of determining the 
    absolute calibration (i.e. the five zero-points) of the SDSS system, choosing 
    instead to have it agree with the published SDSS magnitudes. In particular, 
    any corrections to put the SDSS system on to the AB system also apply in 
    our case.    
\end{itemize}

We conclude by discussing how to extend the program presented here to the next generation of 
imaging surveys. Our starting point will be the second distinction made in Sec.~\ref{sec:intro} -
separating the telescope and the atmosphere explicitly in the calibrations. It is relatively 
straightforward to adapt the algorithm presented here to use high precision measurements of the 
telescope response functions as a starting point; these would be analogous to the priors 
already considered here.

Understanding atmospheric variations is an important step towards 
1\% photometry; unmodeled variations are 
responsible for almost all our calibration error budget. These transparency variations are
dominated by three well-studied processes \citep{1975ApJ...197..593H}: Rayleigh scattering, 
molecular absorption by ozone (dominant in the UV) and water vapor (dominant in the red and IR),
and aerosol scattering. Of these, Rayleigh scattering is best understood, and is well
determined by the local atmospheric pressure. While absorption and aerosol scattering are well
understood in an average sense, their time variation is significant. Tracking these 
would therefore require continuous monitoring of the atmosphere, plus detailed
atmospheric models. The payback for doing so would be a dramatic reduction in calibration errors. 

The algorithm we propose in this paper demonstrates that 1\% relative photometry is 
achievable by the current generation of wide field imaging surveys. The challenge for the 
next generation of surveys is to break through the 1\% barrier.  

NP is supported by a NASA Hubble Fellowship HST-HF-01200.

Funding for the SDSS and SDSS-II has been provided by the Alfred P. Sloan Foundation, the Participating
Institutions, the National Science Foundation, the U.S. Department of Energy, the National Aeronautics
and Space Administration, the Japanese Monbukagakusho, the Max Planck Society, and the Higher Education
Funding Council for England. The SDSS Web Site is http://www.sdss.org/.

The SDSS is managed by the Astrophysical Research Consortium for the Participating Institutions. The
Participating Institutions are the American Museum of Natural History, Astrophysical Institute Potsdam,
University of Basel, Cambridge University, Case Western Reserve University, University of Chicago, Drexel
University, Fermilab, the Institute for Advanced Study, the Japan Participation Group, Johns Hopkins
University, the Joint Institute for Nuclear Astrophysics, the Kavli Institute for Particle Astrophysics
and Cosmology, the Korean Scientist Group, the Chinese Academy of Sciences (LAMOST), Los Alamos National
Laboratory, the Max-Planck-Institute for Astronomy (MPIA), the Max-Planck-Institute for Astrophysics
(MPA), New Mexico State University, Ohio State University, University of Pittsburgh, University of
Portsmouth, Princeton University, the United States Naval Observatory, and the University of Washington.

\bibliography{apjmnemonic,biblio,preprints}
\bibliographystyle{apj}

\end{document}